\documentclass[aps,prd,showpacs,showkeys,nofootinbib,preprintnumbers,onecolumn]{revtex4}
\usepackage{bm}
\usepackage{latexsym}
\usepackage{dcolumn}
\usepackage{amsfonts,amssymb}
\usepackage{graphicx,epsfig}
\usepackage{psfrag}
\usepackage{amsmath}

\def\beq{\begin{equation}}
\def\eeq{\end{equation}} 
\def\br{\begin{eqnarray}}
\def\er{\end{eqnarray}}
\def\benu{\begin{enumerate}}
\def\eenu{\end{enumerate}}
\def\nn{\nonumber} 
\def\pa{{\partial}}
\def\l{\left}
\def\r{\right}

\def\eq#1{{Eq.~(\ref{#1})}}

\begin{document}

%%%%%%%%%%%%%%%%%%%%%%%%%%%%%%%%%%%%%%%%%%%%%%%%%%%%%%%%%%%%%%%%%%%%%%

\title[Reheating in tachyonic inflation and the effects on the
large scale perturbations]{Reheating in tachyonic inflationary 
models:~Effects on the large scale curvature perturbations}
%%%%%%%%%%%%%%%%%%%%%%%%%%%%%%%%%%%%%%%%%%%%%%%%%%%%%%%%%%%%%%%%%%%%%%%%%%%%%%%
\author{Rajeev Kumar Jain$^{1}$\footnote{Present address:~D\'epartement 
de Physique Th\'eorique and Center for Astroparticle Physics,
Universit\'e de Gen\`eve, 24~Quai Ernest-Ansermet, CH--1211 Gen\`eve 4,
Switzerland. E-mail:~rajeev.jain@unige.ch}}
\author{Pravabati Chingangbam$^{2}$\footnote{Present address:~Indian 
Institute of Astrophysics, II Block, Koramangala, Bangalore 560034, India.
E-mail:~prava@iiap.res.in}}
\author{L.~Sriramkumar$^{1}$\footnote{Present address:~Department of 
Physics, Indian Institute of Technology Madras, Chennai 600036, India.
E-mail:~sriram@physics.iitm.ac.in}}
%%%%%%%%%%%%%%%%%%%%%%%%%%%%%%%%%%%%%%%%%%%%%%%%%%%%%%%%%%%%%%%%%%%%%%%%%%%%%%%
\affiliation{$^{1}$Harish-Chandra Research Institute, Chhatnag Road,
Jhunsi, Allahabad~211~019, India.\\
$^{2}$Korea Institute for Advanced Study, 207--43 Cheongnyangni 2-dong,
Dongdaemun-gu, Seoul 130-722, Republic of Korea.}
%%%%%%%%%%%%%%%%%%%%%%%%%%%%%%%%%%%%%%%%%%%%%%%%%%%%%%%%%%%%%%%%%%%%%%%%%%%%%%%
\begin{abstract}
We investigate the problem of perturbative reheating and its effects on
the evolution of the curvature perturbations in tachyonic inflationary
models. 
We derive the equations governing the evolution of the scalar 
perturbations for a system consisting of a tachyon and a perfect fluid.
Assuming the perfect fluid to be radiation, we solve the coupled equations
for the system numerically and study the evolution of the perturbations
from the sub-Hubble to the super-Hubble scales. 
In particular, we analyze the effects of the transition from tachyon 
driven inflation to the radiation dominated epoch on the evolution of 
the large scale curvature and non-adiabatic pressure perturbations. 
We consider two different potentials to describe the tachyon and study 
the effects of two possible types of decay of the tachyon into radiation. 
We plot the spectrum of curvature perturbations at the end of inflation 
as well as at the early stages of the radiation dominated epoch. 
We find that reheating does not affect the amplitude of the curvature
perturbations in any of these cases.
These results corroborate similar conclusions that have been arrived at
earlier based on the study of the evolution of the perturbations in the
super-Hubble limit. 
We illustrate that, before the transition to the radiation dominated 
epoch, the relative non-adiabatic pressure perturbation between the 
tachyon and radiation decays in a fashion very similar to that of the 
intrinsic entropy perturbation associated with the tachyon.
Moreover, we show that, after the transition, the relative non-adiabatic
pressure perturbation dies down extremely rapidly during the early stages
of the radiation dominated epoch. 
It is these behavior which ensure that the amplitude of the curvature
perturbations remain unaffected during reheating. 
We also discuss the corresponding results for the popular chaotic 
inflation model in the case of the canonical scalar field.
\end{abstract}
\maketitle

%%%%%%%%%%%%%%%%%%%%%%%%%%%%%%%%%%%%%%%%%%%%%%%%%%%%%%%%%%%%%%%%%%%%%%%%%%%%%%%

\section{Effects of inflationary and post-inflationary dynamics
on the evolution of the super-Hubble perturbations}

In the standard inflationary scenario, the scales of cosmological 
interest exit the Hubble radius within the first few (about $8$-$10$) 
$e$-folds of inflation and, hence, are outside the Hubble scale 
during the later epochs.
When comparing the predictions of the inflationary models with the 
Cosmic Microwave Background (CMB) and the large scale structure 
data, it is often assumed that the amplitude of the curvature 
perturbations remain constant at super-Hubble scales.
In such a situation, the observed CMB anisotropies directly determine 
the amplitude and the shape of the perturbation spectrum imprinted on 
the modes when they left the Hubble radius during inflation.
Typically, the amplitude of the perturbations constrain the parameters 
that describe the inflaton potential, while the shape of the spectrum 
limits its form~(see any of the standard texts~\cite{texts} or one of 
the following reviews~\cite{reviews}).

Provided inflation is of the slow roll type for {\it all}\/ of 
the required number of $e$-folds, it is indeed true that the 
amplitude of the curvature perturbations freeze at their value 
at Hubble exit.
However, if there is a period of deviation from slow roll inflation, 
then the asymptotic (i.e. the extreme super-Hubble) amplitude of the 
modes that leave the Hubble radius just before the deviation are 
enhanced when compared to their value at Hubble exit~\cite{leach-2001}.
While modes that leave well before the deviation remain unaffected, 
it is found that there exists an intermediate range of modes whose 
amplitudes are actually suppressed at super-Hubble 
scales~\cite{rajeev-2007,bugaev-2008}. 
Depending on the form of the departure from slow roll, these effects 
lead to certain features in the scalar perturbation spectrum. 
If such deviations occur either during or soon after the cosmological 
scales leave the Hubble radius, then the CMB observations constrain 
the resulting features rather well (for an inexhaustive list, see 
Ref.~\cite{fips}).
But, at smaller scales, only theoretical tools are currently available 
to restrict the form of the primordial spectrum.
These constraints are essentially based on the number density of 
primordial black holes that are formed towards the end of inflation 
(for recent discussions in this context, see, for example, 
Ref.~\cite{pbhf}).

Since the cosmological scales are well outside the Hubble radius by 
the early stages of inflation, clearly, the {\it shape}\/ of the 
perturbation spectrum on such large scales is indeed unlikely to be 
affected by subsequent dynamics.
But, over the past decade, it has been recognized that post-inflationary 
dynamics can alter the {\it amplitude}\/ of the curvature perturbations 
at super-Hubble scales, in particular, when more than one component of 
matter is present.
Preheating, the curvaton scenario and the modulated reheating 
mechanism are popular examples that illustrate the interesting 
possibilities of post-inflationary dynamics.
Preheating---a mechanism that transfers energy from the inflaton 
to radiation through an explosive production of quanta corresponding 
to an intermediate scalar field---is known to even lead to an 
exponential growth in the amplitude of the super-Hubble 
perturbations (for the earlier discussions, see Ref.~\cite{ph-ee} and, 
for more recent efforts, see, for instance, Ref.~\cite{ph-mre}).
In the curvaton scenario, while the inflationary epoch still 
remains the source of the perturbations, these perturbations 
are amplified after inflation due to the presence of entropic 
perturbations (see, for example, Ref.~\cite{cs}). 
The modulated reheating scenario is an extreme case wherein 
inflation is essentially required only to resolve the horizon 
problem, whereas the perturbations are generated due to an 
inhomogeneous decay rate when the energy is being transferred 
from the inflaton to radiation through other
fields~\cite{matarrese-2003,mazumdar-2003,mrh}.
These different alternatives indicate that the post-inflationary 
evolution of the large scale curvature perturbations can be highly 
model dependent and, therefore, requires a careful and systematic 
study.
Evidently, in these scenarios, the effects of post-inflationary 
dynamics have to be taken into account when constraining the
inflationary models using the CMB observations (in this context,
see, for instance, Ref.~\cite{dimopoulos-2004}).

With these motivations in mind, in this paper, we investigate the 
problem of the more conventional (perturbative) reheating 
scenario~\cite{rh,dimarco-2007,cerioni-2008}, and its effects on 
the evolution of the curvature perturbations in tachyonic inflationary 
models (for the original discussions on the tachyon, see 
Refs.~\cite{tachyon,sen-2002}; for efforts on treating the tachyon 
as an inflaton, see, for instance, 
Refs.~\cite{sami-2002,steer-2004,prava-2005}; for discussions on 
reheating in such inflationary models, see 
Refs.~\cite{campuzano-2005-2006,cardenas-2006}). 
We shall consider two types of potentials to describe the tachyon 
and construct scenarios of transition from inflation to radiation 
domination for the following two possible cases of the decay rate 
$\Gamma$ of the tachyon into radiation: (i)~a constant, and 
(ii)~dependent on the tachyon.
We solve the coupled equations for the system numerically and 
study the evolution of the perturbations from the sub-Hubble to 
the super-Hubble scales. 
Importantly, we shall consider the effects of the transition from 
inflation to the radiation dominated epoch on the evolution of the 
large scale (i.e. those that correspond to cosmological scales 
today) curvature and non-adiabatic  pressure (i.e. the intrinsic 
entropy as well as the relative entropy or the isocurvature) 
perturbations. 
We shall evaluate the spectrum of curvature perturbations at the 
end of inflation as well as at the early stages of the radiation 
dominated epoch.
As we shall illustrate, reheating does not affect the amplitude of 
the curvature perturbations in any of these cases.
We shall also show that, before the transition to the radiation 
dominated epoch, the relative non-adiabatic pressure perturbation 
between the tachyon and radiation decays in a fashion very similar 
to that of the intrinsic entropy perturbation associated with the 
tachyon. 
Moreover, we demonstrate that, after the transition, the 
relative non-adiabatic pressure perturbation dies down extremely 
rapidly during the early stages of the radiation dominated epoch. 
It is these behavior which ensure that the amplitude of the 
curvature perturbations remains unaffected during reheating.
Our results corroborate similar conclusions that have been arrived 
at earlier in the literature based on the study of the evolution of 
the perturbations in the super-Hubble limit~\cite{matarrese-2003}.
We shall also discuss similar effects in the case of the canonical 
scalar field.

A few clarifying remarks are in order at this stage of the discussion.
We should stress that the set up we are considering is the standard
cold inflationary scenario, followed by an epoch of reheating achieved
by the standard method of introducing a coarse-grained decay rate 
in the equation of motion describing the 
inflaton~\cite{rh,dimarco-2007}.
Recently, an analysis somewhat similar to what we shall consider 
here has been studied in the context of warm inflationary scenarios
involving tachyonic fields~\cite{wti}.
Though there can exist similarities in the form of the equations in 
the cold and the warm inflationary scenarios, the physics in these 
two scenarios are rather different (for a recent discussion, see,
for example, Ref.~\cite{wi}).
Moreover, to analyze the effects of reheating on the large scale
perturbations, one could have possibly studied the evolution of 
the perturbations in the super-Hubble limit, say, the first order 
(in time) differential equations usually considered in the 
literature (cf. Refs.~\cite{matarrese-2003,mazumdar-2003,malik-2003}).
However, some concerns have been raised that the coupling between
the inflaton and radiation may affect the amplitude as well as the 
scalar spectral index in certain 
situations~\cite{dimarco-2007,cerioni-2008}.
Also, the modified background dynamics preceding reheating can 
effect the extent of small scale primordial black holes that are 
formed towards the end of inflation.
If we are to address such issues, it requires that we study 
the evolution of the perturbations from the sub-Hubble to the 
super-Hubble scales. 
We shall comment further on some of these points in the concluding 
section.

The remainder of this paper is organized as follows.
In the following section, we shall summarize the essential 
background equations for the system consisting of the tachyon and a 
perfect fluid and set up scenarios of transition from inflation to 
radiation domination.
In Sec.~\ref{sec:sp}, we shall obtain the equations describing the 
scalar perturbations for the system of the tachyon and a perfect 
fluid.
In Sec.~\ref{sec:ep}, we shall evolve the coupled system of
equations describing the perturbations and study the effects 
of the transition from inflation to the radiation dominated 
epoch on the super-Hubble curvature perturbations. 
We shall plot the spectrum of curvature perturbations at the end 
of inflation as well as at the early stages of the radiation 
dominated epoch. 
We shall also explicitly illustrate that the intrinsic entropy 
perturbation associated with the tachyon and the relative 
non-adiabatic pressure perturbation between the tachyon and radiation 
decay in a similar fashion before and after the transition to the
radiation dominated epoch.
It is the behavior of these non-adiabatic pressure perturbations which 
ensure that the amplitude of the curvature perturbations remains 
unaffected during reheating.
Finally, in Sec.~\ref{sec:d}, we conclude with a summary and a discussion 
on the results we have obtained.
In the appendix, we shall discuss the corresponding results for the 
chaotic inflation model in the case of the canonical scalar field.

The notations and conventions we shall adopt are as follows.
We shall set $\hbar=c=1$, but shall display $G$ explicitly, and
define the Planck mass to be $M_{_{\rm Pl}}=\l(8\, \pi\, G\r)^{-1/2}$.
We shall work with the metric signature of $(+,-,-,-)$.
We shall express the various quantities in terms of the cosmic 
time~$t$, and we shall denote differentiation with respect to~$t$ 
by an overdot.
While we shall use the subscript $i$ to denote the spatial components 
of, say, the momentum flux associated with the matter fields, the 
subscripts $\alpha$ and $\beta$ shall refer to the two components 
of matter.

%%%%%%%%%%%%%%%%%%%%%%%%%%%%%%%%%%%%%%%%%%%%%%%%%%%%%%%%%%%%%%%%%%%%%%%%%%%%%%%

\section{The transition from tachyon driven inflation to the radiation
dominated epoch}\label{sec:be}

In this section, we shall discuss the background equations describing 
the evolution of the tachyon that is interacting with a perfect fluid.
Assuming the perfect fluid to be radiation, we shall construct 
scenarios of transition from tachyon driven inflation to radiation 
domination for the following two possible types of the decay rate 
$\Gamma$ [cf. \eq{eq:QF} below] of the inflaton into radiation: 
(i)~a constant, and (ii)~dependent on the tachyon~\cite{matarrese-2003}. 

%%%%%%%%%%%%%%%%%%%%%%%%%%%%%%%%%%%%%%%%%%%%%%%%%%%%%%%%%%%%%%%%%%%%%%%%%%%%%%%

\subsection{Background equations in the presence of
interacting components}

Consider a $(3 + 1)$-dimensional, spatially flat, smooth, expanding,
Friedmann universe described by the line element
\beq
{\rm d}s^2 = {\rm d}t^2-a^{2}(t)\; {\rm d}{\bf x}^2,\label{eq:frwle}
\eeq
where $t$~denotes the cosmic time and $a(t)$~is the scale factor.
Let $\rho$ and $p$ denote the total energy density and the total 
pressure of a system consisting of multiple components of fields 
and fluids that are driving the expansion.
Then, the Einstein's equations corresponding to the above line-element 
lead to the following Friedmann equations for the scale factor~$a(t)$:
\beq
H^{2}= \l(\frac{8\,\pi\, G}{3}\r)\; {\rho}\qquad{\rm and}\qquad
{\dot H}= -\l(4\,\pi\, G\r) \l(\rho + p\r),\label{eq:fe}
\eeq
where $H=\l({\dot a}/a\r)$ is the Hubble parameter. 
Also, the conservation of the total energy of the system leads 
to the continuity equation
\beq
{\dot \rho} + 3\,H \l(\rho + p\r) = 0.\label{eq:tec}
\eeq

The total energy density and the total pressure of the system can be
expressed as the sum of the energy density $\rho_{_{\alpha}}$ and the
pressure $p_{_{\alpha}}$ of the individual components as follows:
\beq
\rho = \sum_\alpha\, \rho_{_{\alpha}}\qquad{\rm and}\qquad
p = \sum_\alpha\, p_{_{\alpha}}.
\eeq
If the different components of fields and fluids do not interact,
then, in addition to the total energy density, the energy density
of the individual components will be conserved as well.
Hence, in such a situation, the energy density $\rho_{_{\alpha}}$ 
of each component will individually satisfy the continuity equation
\beq
\dot{\rho_{_{\alpha}}}
+3\,H \l(\rho_{_{\alpha}} + p_{_{\alpha}}\r) = 0.
\eeq
On the other hand, when the different components interact, the 
continuity equation for the individual components can be expressed as 
(see, for instance, Refs.~\cite{matarrese-2003,malik-2001,malik-2005})
\beq
\dot{\rho_{_{\alpha}}}+3\,H \l(\rho_{_{\alpha}} + p_{_{\alpha}}\r)
= Q_{_{\alpha}},\label{eq:ceic}
\eeq
where $Q_{_{\alpha}}$ denotes the rate at which energy density is 
transferred to the component~$\alpha$ from the other components. 
The conservation of the energy of the complete system then leads 
to the following constraint on the total rate of transfer of the 
energy densities:
\beq
\sum_\alpha\, Q_{_{\alpha}}=0.\label{eq:c}
\eeq

%%%%%%%%%%%%%%%%%%%%%%%%%%%%%%%%%%%%%%%%%%%%%%%%%%%%%%%%%%%%%%%%%%%%%%%%%%%%%%%

\subsection{The case of the tachyon and a perfect fluid}

Let us now consider the system of a tachyon (which we shall denote 
as~$T$) that is interacting with a perfect fluid (referred to, 
hereafter, as~$F$), so that $\alpha=(T,F)$.
We shall assume that the rate at which energy density is transferred 
to the perfect fluid, i.e. $Q_{_{F}}$, is given by~\cite{cardenas-2006}
\beq
Q_{_{F}}= \Gamma\; \dot{T}^2\, \rho_{_{T}},\label{eq:QF}
\eeq
where $\rho_{_{T}}$ is the energy density of the tachyon. 
Such a transfer of energy is assumed to describe---albeit, in a 
course grained fashion---the perturbative decay of the tachyon 
into particles that constitute the perfect fluid.
The quantity~$\Gamma$ represents the corresponding decay rate and, 
as we shall discuss below, it can either be a constant or depend on
the tachyon~\cite{matarrese-2003}.

For the above $Q_{_{F}}$, the continuity equation~(\ref{eq:ceic})
corresponding to the perfect fluid is given by
\beq
{\dot \rho}_{_{F}} + 3\,H \l(1 + w_{_{F}}\r)\, \rho_{_{F}} 
= \Gamma\; \dot{T}^2\, \rho_{_{T}},\label{eq:ceF}
\eeq
where $w_{_{F}}=(p_{_{F}}/\rho_{_{F}})$ is the equation of state
parameter describing the perfect fluid, which we shall assume to
be a constant.
Also, it is evident from Eq.~(\ref{eq:c}) that
\beq
Q_{_{T}}=-Q_{_{F}} = -\Gamma\; \dot{T}^2\, \rho_{_{T}}.
\eeq
Therefore, the continuity equation for the tachyon energy 
density~$\rho_{_{T}}$ reduces to
\beq
{\dot \rho}_{_{T}} + 3\,H \l(\rho_{_{T}} + p_{_{T}}\r)
= -\Gamma\; \dot{T}^2\, \rho_{_{T}},
\label{eq:ceT}
\eeq
where $p_{_{T}}$ is the pressure associated with the tachyon.
Given a potential $V(T)$ describing the tachyon, the corresponding 
energy density $\rho_{_{T}}$ and pressure $p_{_{T}}$ are given by 
(see, for instance, Refs.~\cite{rajeev-2007,steer-2004,prava-2005})
\beq
\rho_{_{T}} = \l(\frac{V(T)}{\sqrt{1 - {\dot T}^2}}\r)
\qquad{\rm and}\qquad
p_{_{T}} = -V(T)\; \sqrt{1 - {\dot T}^2}\,.
\eeq
On substituting these two expressions in the continuity 
equation~(\ref{eq:ceT}), we arrive at the following equation 
of motion for the tachyon~$T$~\cite{cardenas-2006}:
\beq
\l(\frac{\ddot T}{1-{\dot T}^2}\r)
+3\, H\; {\dot T}+\Gamma\; {\dot T}
+\l(\frac{V_{_{T}}}{V}\r)=0, 
\label{eq:emT}
\eeq
where $V_{_{T}}\equiv \l({\rm d}V/{\rm d}T\r)$. 

Before we proceed, the following clarification on the choice of 
$Q_{_{F}}$ in \eq{eq:QF} is in order at this stage of the 
discussion.
Recall that, in a situation wherein the canonical scalar field, 
say~$\phi$, drives inflation, it is common to introduce a 
$(\Gamma\, {\dot \phi})$ term in the field equation to describe 
the perturbative decay of the inflaton [see 
Refs.~\cite{texts,reviews,rh,dimarco-2007,cerioni-2008}, also see 
\eq{eq:emcsf}].
Motivated by the canonical case, the choice of $Q_{_{F}}$ in 
\eq{eq:QF} has been specifically made so as to lead to 
the $(\Gamma\, {\dot T})$ term in the equation of 
motion~(\ref{eq:emT}) for the tachyon~\cite{cardenas-2006}.
Needless to add, the choice~(\ref{eq:QF}) is but one of the many
possibilities that can help in achieving reheating at the end of 
inflation. 

%%%%%%%%%%%%%%%%%%%%%%%%%%%%%%%%%%%%%%%%%%%%%%%%%%%%%%%%%%%%%%%%%%%%%%%%%%

\subsection{Tachyonic inflationary models, different possible 
$\Gamma$ and reheating}\label{sec:models}

In this sub-section, assuming the perfect fluid to be radiation 
with $w_{_{F}}=(1/3)$, we shall construct specific scenarios of 
transition from tachyon driven inflation to an epoch of radiation 
domination.
 
We shall consider two different types of potentials in order to 
describe the tachyon.
\begin{itemize}
\item
The first potential we shall consider is given by
\beq
V_{1}(T)=\l(\frac{\lambda}{{\rm cosh}\,(T/T_{_{0}})}\r),
\eeq
a potential that is well motivated from the string theory 
perspective~\cite{sen-2002}. 
\item
Our second choice will be the following phenomenologically 
motivated power law potential that has been considered
earlier in the literature (see, for example, 
Ref.~\cite{steer-2004}):
\beq
V_{2}(T)=\l(\frac{\lambda}{1+(T/T_{_{0}})^{4}}\r).
\eeq
\end{itemize}
In order to achieve the necessary amount of inflation and 
the correct amplitude for the scalar perturbations, suitable 
values for the two parameters $\lambda$ and $T_{_{0}}$ that 
describe the above potentials can be arrived at as follows.
Firstly, one finds that, in these potentials, inflation 
typically occurs around $T\simeq T_{_{0}}$ corresponding to 
an energy scale of about~$\lambda^{1/4}$.
Secondly, it turns out that, the quantity $(\lambda\, T_{_{0}}^2/
M_{_{\rm Pl}}^2)$ has to be much larger than unity for the potential
slow roll parameters to be small and thereby ensure that, at least, 
$60$ $e$-folds of inflation takes place. 
Therefore, one first chooses a sufficiently large value of 
$(\lambda\, T_{_{0}}^2/M_{_{\rm Pl}}^2)$ by hand in order to guarantee 
slow roll. The COBE normalization condition for the scalar perturbations 
then provides the second constraint, thereby determining the values 
of both the parameters~$\lambda$ and $T_{_{0}}$~\cite{steer-2004}.

In the absence of the fluid, we find that the two potentials 
$V_{1}(T)$ and $V_{2}(T)$ above allow about $60$ $e$-folds of 
slow roll inflation and lead to the correct COBE amplitude for 
the choice of the parameters and initial conditions listed in 
Table~\ref{tab:beiaf}.
\begin{table}[!htb]
\vskip 4 pt
\begin{center}
\begin{tabular}{|c|c|c|c|c|}
\hline
Potential & $\lambda$    & $T_{_{0}}$   & 
$(T_{i}/T_{_{0}})$ & ${\dot T}_{i}$ \\
\hline
$V_{1}(T)$  & $10^{-6}$ & $10^5$  & $5.81$    
& $0.1$\\
\hline
$V_{2}(T)$  & $10^{-5.15}$ &  $3.7\times 10^4$ & 
$4.56$ & $0.1$\\
\hline
\end{tabular}
\caption{\label{tab:beiaf}Achieving inflation:~The values for 
the parameters of the tachyon potential and the initial conditions 
for the inflaton (the initial value~$T_{i}$ of the tachyon 
and its initial velocity~${\dot T}_{i}$) that we work with.
These choices for the parameters and initial conditions lead 
to the required $60$ $e$-folds of slow roll inflation and the 
observed amplitude for the scalar perturbations.}
\end{center}
\end{table}
Also, in such a situation, it is known that, at the end 
of inflation, the tachyon leads to an epoch of dust like 
behavior~\cite{sami-2002,steer-2004}, thereby making a 
radiation dominated epoch difficult to achieve. 
However, as we shall illustrate below, when the tachyon is 
interacting with the fluid, we can ensure that a transition 
from inflation to the radiation dominated regime occurs with 
a suitable choice of the amplitude for the decay rate~$\Gamma$.

\begin{figure}[!htb]
\begin{center}
\vskip 15pt\hskip 15pt
\resizebox{200pt}{130pt}{\includegraphics{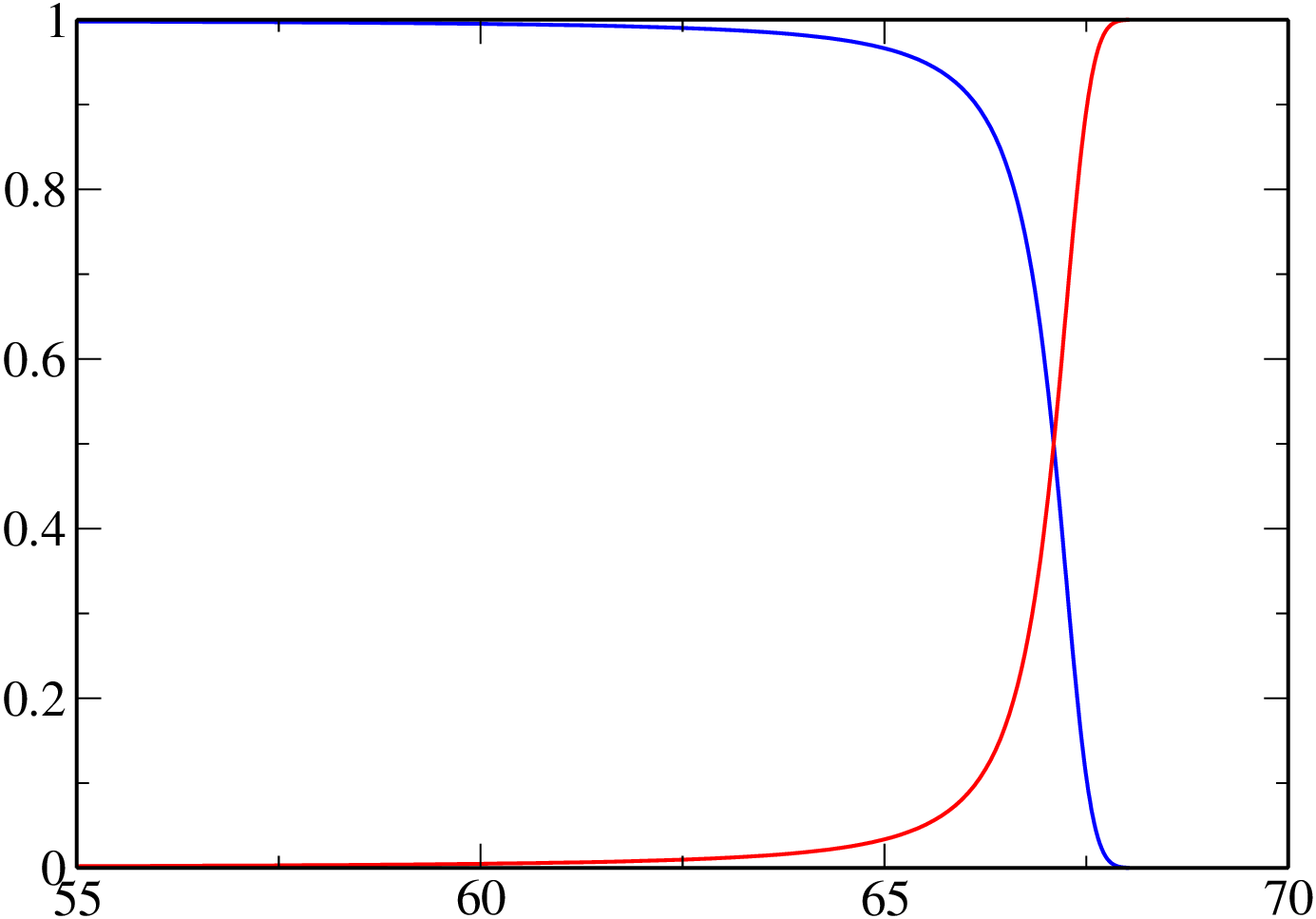}}
\hskip 25pt
\resizebox{200pt}{130pt}{\includegraphics{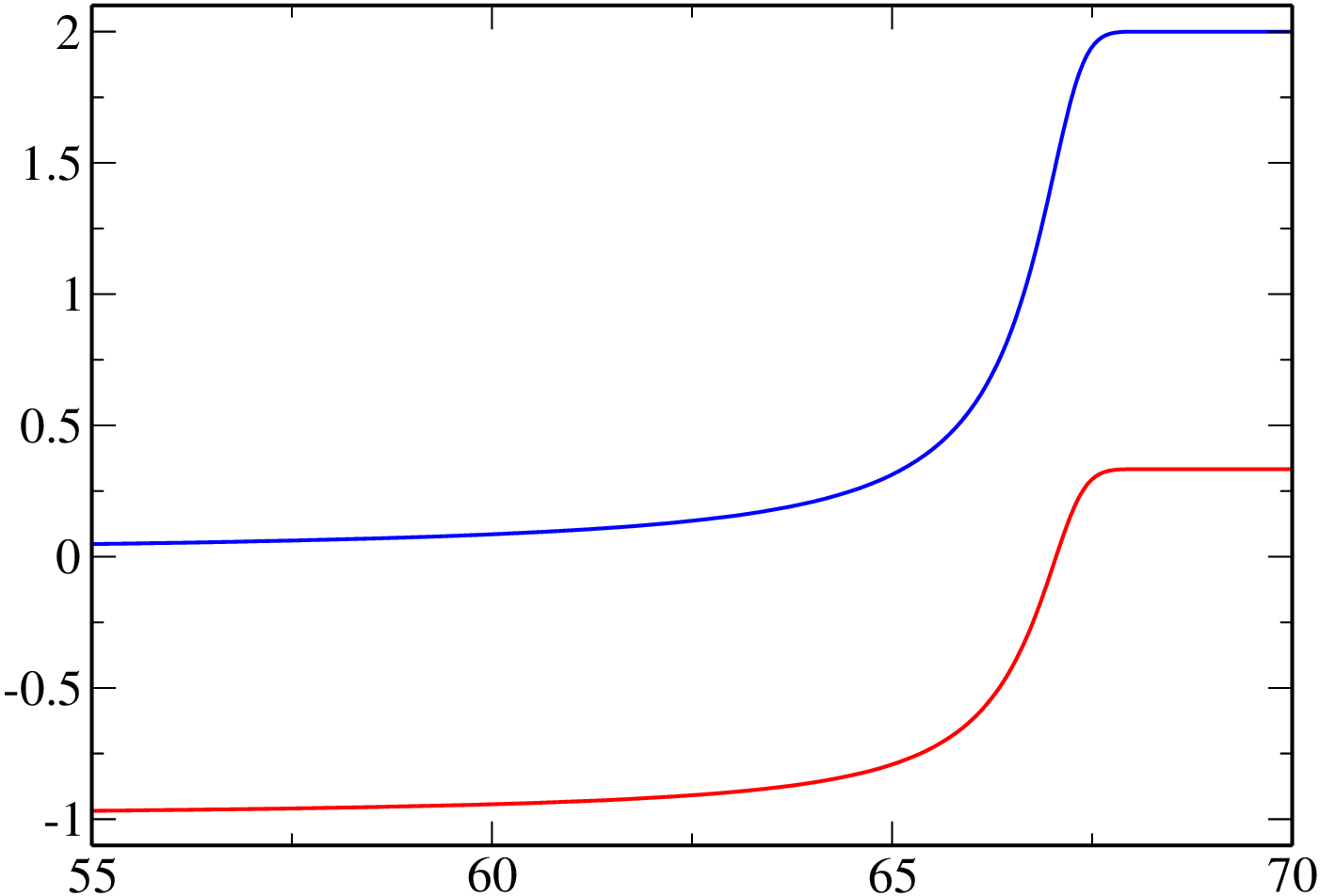}}
\vskip -100 true pt 
\hskip -220 true pt $\Omega_{_{T}}, \Omega_{_{\gamma}}$
\hskip 210 true pt $\epsilon,w$
\vskip 105 pt\hskip 15pt
\resizebox{200pt}{130pt}{\includegraphics{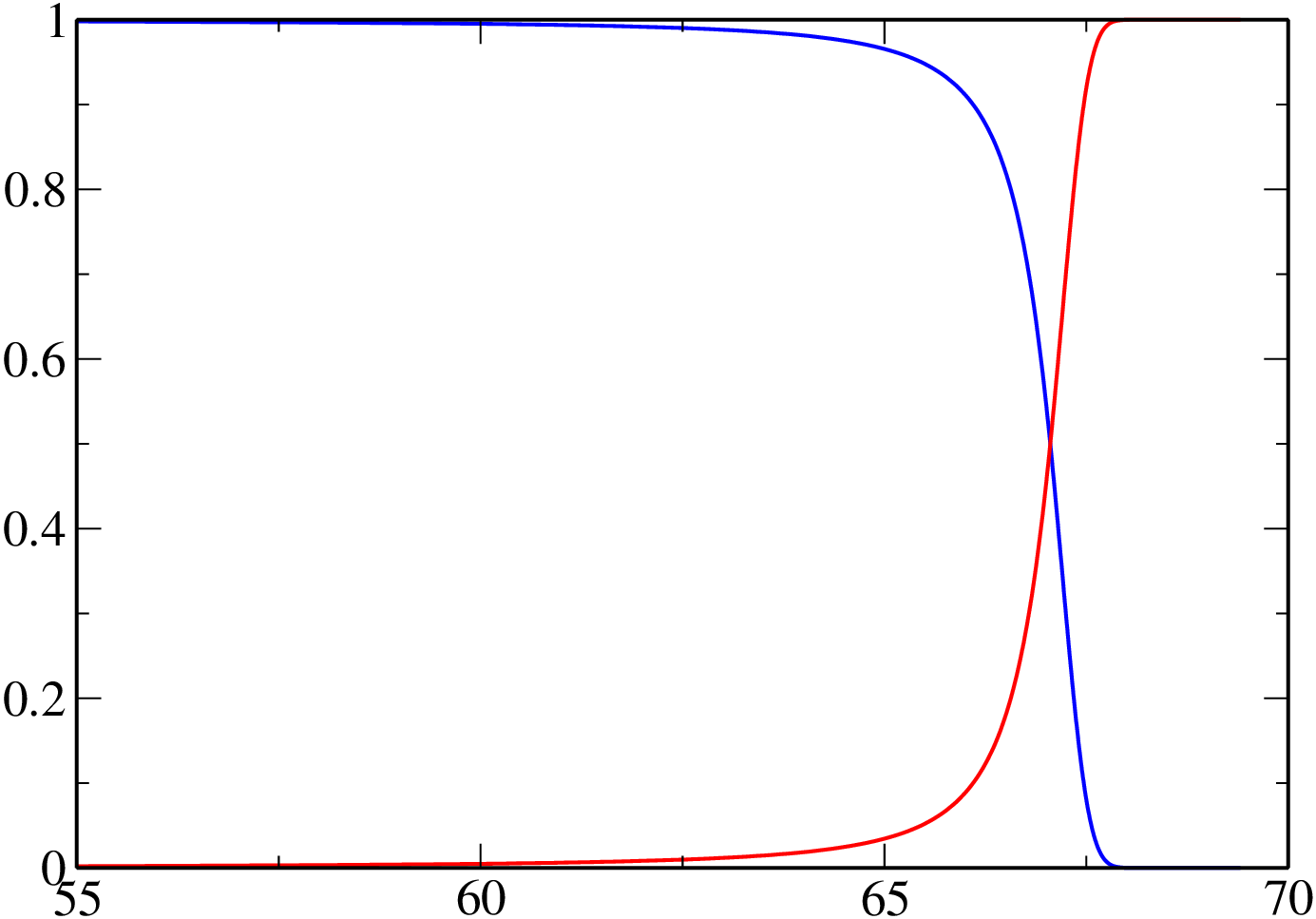}}
\hskip 25pt
\resizebox{200pt}{130pt}{\includegraphics{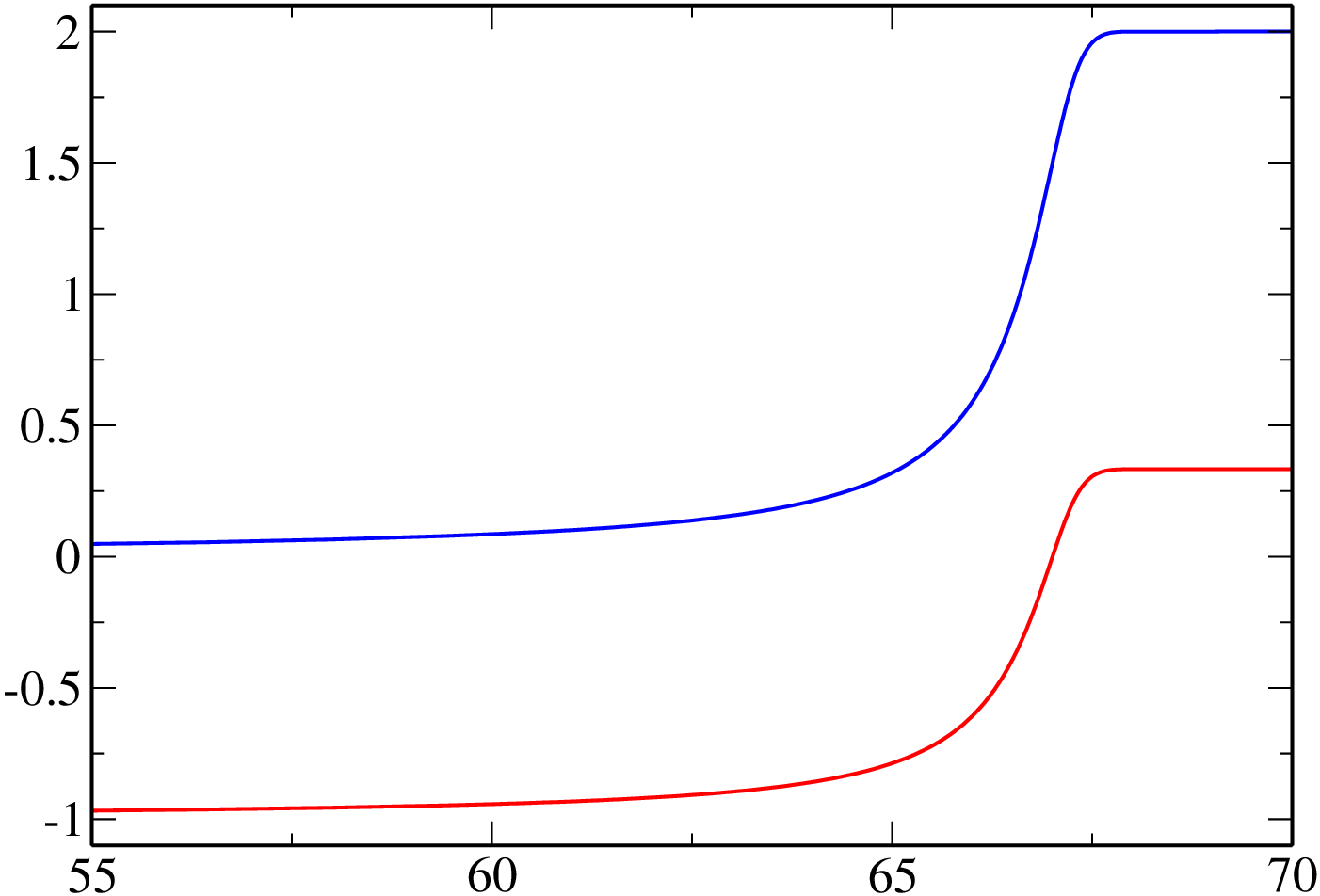}}
\vskip -105 true pt 
\hskip -220 true pt $\Omega_{_{T}}, \Omega_{_{\gamma}}$
\hskip 210 true pt $\epsilon,w$
\vskip 95 pt
\hskip 22 true pt $N$ \hskip 215 true pt $N$
\vskip 5pt
\caption{\label{fig:eewdbif}In the left column, the evolution of the 
quantities $\Omega_{_{T}}$ (in blue) and $\Omega_{_{\gamma}}$ (in red)
has been plotted as a function of the number of $e$-folds~$N$ for the 
tachyon potential $V_{1}(T)$ and the two cases of $\Gamma$, viz. 
$\Gamma_{1}={\rm constant}$ (on top) and $\Gamma_{2} = \Gamma(T)$ (at 
the bottom), with $\Gamma(T)$ given by Eq.~(\ref{eq:gammaT}).
Similarly, in the right column, the evolution of the first Hubble slow 
roll parameter~$\epsilon$ (in blue) and the equation of state parameter 
of the entire system~$w$ (in red) has been plotted as a function of the 
number of $e$-folds for the same potential and the different cases of 
$\Gamma$, as indicated above. 
All these plots correspond to the values for the various parameters and 
the initial conditions that we have listed in Tables~\ref{tab:beiaf}
and~\ref{tab:beipf}.
The figures clearly illustrate the transfer of energy from the inflaton
to radiation. 
(Note that $\epsilon=2$ during the radiation dominated epoch.)
We have chosen the various parameters in such a fashion that there is 
a sufficiently rapid transition from inflation to radiation domination 
without any intermediate regime, as suggested by the evolution of 
$\epsilon$ and $w$.
We find that a very similar behavior occurs for the potential $V_{2}(T)$.}
\end{center}
\end{figure}

As we mentioned before, we shall consider the following two 
possible choices for the decay rate $\Gamma$:~(i)~$\Gamma_{1}
={\cal A}={\rm constant}$, and (ii)~$\Gamma_{2} = \Gamma(T)$.
We shall work with the following specific form of $\Gamma(T)$:
\beq 
\Gamma_{2} = \Gamma(T) 
= {\cal A}\; \biggl(1-{\cal B}\; 
{\rm tanh}\,\l[\l(T-{\cal C}\r)/{\cal D}\r]\biggr),
\label{eq:gammaT}
\eeq
where ${\cal A}$, ${\cal B}$, ${\cal C}$ and ${\cal D}$ are 
constants that we shall choose suitably to achieve the desired 
evolution. (It turns out that the above form of $\Gamma(T)$ prove to be 
convenient in the numerical calculations, helping us 
illustrate the required behavior. We shall comment further on 
this choice in the concluding section.)
We find that a transition from inflation to radiation domination
can be achieved provided the amplitude of $\Gamma$ (viz.~${\cal A}$) 
is chosen to be less than the Hubble parameter $H$ (which is almost a 
constant) during the epoch of slow roll inflation.
\begin{table}[!htb]
\begin{center}
\begin{tabular}{|c|c|c|c|c|c|c|}
\hline
Potential & $\Gamma$ & $\rho_{_{\gamma}}^{i}$ & ${\cal A}$ 
& ${\cal B}$ & ${\cal C}$ & ${\cal D}$\\
\hline
$V_{1}(T)$ & $\Gamma_{1}$ & $10^{-40}$ & $5\, \lambda$ 
& -- & -- & --\\
\cline{2-7}
& $\Gamma_{2}$ & $10^{-40}$ & $5\, \lambda$ 
& $10^{-2}$ & $2\, T_{_{0}}$ & $T_{_{0}}$ \\
\hline
$V_{2}(T)$ & $\Gamma_{1}$ & $10^{-40}$ & $1.43\times 10^{-2}\, 
\lambda$ & -- & -- &--\\
\cline{2-7}
& $\Gamma_{2}$ & $10^{-40}$ & $10^{-2}\, \lambda$ 
& $10^{-2}$& $2\, T_{_{0}}$ & $T_{_{0}}$\\
\hline
\end{tabular}
\caption{\label{tab:beipf}Achieving inflation {\it and}\/
reheating:~The values of the initial energy density of radiation
$\rho_{_{\gamma}}^{i}$ and the parameters describing the decay 
rates that we work with when the interaction of the tachyon with 
radiation is taken into account.
For the tachyon, we use the same values listed in the previous 
table. 
These parameters and initial conditions lead to about 
$60$ $e$-folds of slow roll inflation, the required amplitude 
for the perturbations, and also reheat the universe within a 
couple of $e$-folds after inflation.} 
\end{center}
\end{table} 
In Table~\ref{tab:beipf}, we have listed the values of the initial energy 
density of radiation (viz. $\rho_{_{\gamma}}^{i}$) and the various 
parameters describing the decay rates (for the tachyon, we have used 
the same values listed in Table~\ref{tab:beiaf}) that lead to about $60$ 
$e$-folds of slow roll inflation, the correct scalar amplitude, while 
also reheating the universe within $2$-$3$ $e$-folds after the end of 
inflation.
In the left column of Fig.~\ref{fig:eewdbif}, we have plotted 
the fractional contributions (with respect to the total 
energy density) of the energy densities of the tachyon and 
radiation, i.e. $\Omega_{_{T}}=(\rho_{_{T}}/\rho)$ and 
$\Omega_{_{\gamma}}=(\rho_{_{\gamma}}/\rho)$, as a function 
of the number of $e$-folds $N$ for the potential~$V_{1}(T)$.
It is evident from the figure that the two choices for the decay rate 
$\Gamma$ ensure the completion of reheating (i.e. $\Omega_{_{\gamma}}
\simeq 1$) within a couple of $e$-folds after the end of inflation. 
In the right column of the figure, we have plotted the corresponding 
effective equation of state parameter, viz. $w=(p/\rho)$, and also the 
first Hubble slow roll parameter, viz. $\epsilon=-({\dot H}/H^2)$, of 
the entire system, for the same tachyon potential.
These plots, while confirming that inflation has indeed ended and 
reheating has been realized, also indicate the nature of the composite 
matter during the transition.
We should add that a very similar behavior occurs for the 
potential~$V_{2}(T)$.

%%%%%%%%%%%%%%%%%%%%%%%%%%%%%%%%%%%%%%%%%%%%%%%%%%%%%%%%%%%%%%%%%%%%%%%%%%%%%%

\section{Equations of motion governing the scalar perturbations 
for the system}\label{sec:sp}

In this section, using the first order Einstein's equations and the
equations describing the conservation of the perturbed energy density
of the tachyon and the perfect fluid, we shall arrive at the coupled, 
second order (in time) differential equations governing the scalar 
perturbations for the system consisting of the tachyon and the fluid.
For convenience, we shall work in the uniform curvature gauge (UCG),
a gauge that is also referred to, often, as the spatially flat gauge.

%%%%%%%%%%%%%%%%%%%%%%%%%%%%%%%%%%%%%%%%%%%%%%%%%%%%%%%%%%%%%%%%%%%%%%%%%%%%%%%

\subsection{First order Einstein's equations}

If we take into account the scalar perturbations to the spatially flat 
background metric~(\ref{eq:frwle}), then, in the UCG, the Friedmann 
line-element is given by~\cite{texts,reviews}
\beq
{\rm d}s^2 
= (1+2\, A)\,{\rm d}t^2 - 2\, a\, (\pa_{i} B)\, {\rm d}x_i\, 
{\rm d}t - a^2(t)\; {\rm d}{\bf x}^2,\label{eq:frwlesp}
\eeq
where $A$ and $B$ are functions that describe the two degrees of freedom
associated with the perturbations.
Being a scalar field, the tachyon, evidently, does not possess any 
anisotropic stress at the linear order in the perturbations. 
We shall assume that the perfect fluid does not contain any anisotropic 
stress either.
In such a case, at the first order in the perturbations, Einstein's 
equations in the UCG can be written as~\cite{texts,reviews}
\br
- 3\, H^2 A - \l(\frac{H}{a}\r)\, \nabla^2 B
&=& (4\,\pi\, G)\, \delta \rho,\label{eq:foee00}\\
H \l(\pa_{i}\,A\r)
&=& (4\,\pi\, G) \l(\pa_{i}\,{\delta q}\r),\label{eq:foee0i}\\
H\, {\dot A} + \l(2\,{\dot H} + 3\, H^2\r)\, A
&=& (4\,\pi\, G)\, \delta p,\label{eq:foeeii}
\er
where the quantities $\delta\rho$, $(\pa_{i}\,{\delta q})$, and $\delta p$ 
denote the perturbations in the total energy density, the total momentum 
flux, and the total pressure of the complete system, respectively.
The absence of the anisotropic stress leads to the following 
additional relation between the two functions $A$ and $B$ 
describing the scalar perturbations:
\beq
A+a \l({\dot B}+2\, H\, B\r)=0.\label{eq:foeeij}
\eeq
For the system of our interest, viz. that of the tachyon and a perfect
fluid described by a constant equation of state parameter $w_{_{F}}$, 
the quantities $\delta\rho$, ${\delta q}$ and $\delta p$ are given by
\br
\delta \rho
\ =\ \l(\delta \rho_{_{T}} +\delta\rho_{_{F}}\r)
&=&\l(\frac{V_{_{T}}\, {\delta T}}{\sqrt{1-{\dot T}^{2}}}\r)\,
+\l(\frac{V\, \dot T}{\l(1-{\dot T}^2\r)^{3/2}}\r)\, 
\l({\dot {\delta T}}- A\, {\dot T}\r)
+\delta \rho_{_{F}},\label{eq:deltarho}\\
\delta q 
\ =\ \l(\delta q_{_{T}} +\delta q_{_{F}}\r)
&=& \l(\frac{V\, {\dot T}\,\delta T}
{\sqrt{1-{\dot T}^2}}\r) - \psi_{_{F}},\label{eq:deltaq}\\
\delta p 
\ =\ \l(\delta p_{_{T}} +\delta p_{_{F}}\r)
&=& -\l(V_{_{T}}\, \delta T\, \sqrt{1-{\dot T}^2}\r)
+\l(\frac{V\, {\dot T}}{\sqrt{1-{\dot T}^{2}}}\r)\, 
\l({\dot {\delta T}}- A\, {\dot T}\r)
+w_{_{F}} \delta\rho_{_{F}},\label{eq:deltap}
\er
where $\delta T$ denotes the perturbation in the tachyon and 
$\psi_{_{F}}$ is proportional to the potential that determines 
the three (i.e. the spatial) velocity of the perfect 
fluid\footnote{Usually, for a perfect fluid, the quantity 
$\delta q_{_{F}}$ is written as $[(\rho_{_{F}}+p_{_{F}})\, 
\chi_{_{F}}]$, where $\chi_{_{F}}$ is the potential whose spatial 
gradient describes the three velocity of the fluid (see, 
for instance, Refs.~\cite{texts,reviews}). 
For convenience, we have instead defined the entire quantity 
$[(\rho_{_{F}}+p_{_{F}})\; \chi_{_{F}}]$ as 
$-\psi_{_{F}}$~\cite{dimarco-2007}.
As we shall point out later, we have ensured that suitable sub-Hubble
initial conditions are imposed on $\psi_{_{F}}$.}.  

%%%%%%%%%%%%%%%%%%%%%%%%%%%%%%%%%%%%%%%%%%%%%%%%%%%%%%%%%%%%%%%%%%%%%%%%%

\subsection{Equations of motion describing the perturbed matter 
components}\label{sec:perteqns}

In the UCG, at the first order in the perturbations, the equation 
describing the conservation of the energy density of the component 
$\alpha$ is given by~\cite{reviews,malik-2003,malik-2001,malik-2005}
\beq
\dot{\delta\rho}_{_{\alpha}} + 3\,H\;(\delta\rho_{_{\alpha}}
+\delta p_{_{\alpha}}) - \l(\frac{1}{a^{2}}\r) \nabla^2
\l[{\delta q}_{_{\alpha}} + (\rho_{_{\alpha}}+ p_{_{\alpha}})\, a\,B\r] 
- Q_{_{\alpha}}\,A - \delta Q_{_{\alpha}}=0,\label{eq:pece}
\eeq
where the quantities $\delta \rho_{_{\alpha}}$, ${\delta q}_{_{\alpha}}$
and $\delta p_{_{\alpha}}$ denote the perturbations at the linear order 
in the energy density, the momentum flux, and the pressure of the 
particular component $\alpha$, while $Q_{_{\alpha}}$ and 
$\delta Q_{_{\alpha}}$ indicate the rate at which the energy is 
transferred to the component~$\alpha$ and the perturbation in the 
rate, respectively.
The equation of motion that governs the perturbation $\delta T$ 
in the tachyon can now be obtained upon using the 
expressions~(\ref{eq:deltarho})--(\ref{eq:deltap}) for 
$\delta \rho_{_{T}}$, $\delta q_{_{T}}$ and $\delta p_{_{T}}$ in 
above equation for the conservation of the perturbed energy density. 
We find that the equation of motion describing the perturbation 
$\delta T$ is given by
\br
\l(\frac{\ddot{\delta T}}{1-\dot T^2}\r)
+ \l[\l(3\,H+\Gamma\r) \l(1-3\,{\dot T}^2\r)
-2\, {\dot T} \l(\frac{V_{_{T}}}{V}\r)\r]
\l(\frac{\dot{\delta T}}{1-{\dot T}^2}\r)
+\l[\l(\frac{d^2{\rm ln}\,V}{dT^2}\r)
-\l(\frac{1}{a^2}\r) \nabla^{2}\r] {\delta T}\nn\\
-\l(\frac{{\dot T}\, {\dot A}}{1-{\dot T}^2}\r)
+ \l[6\, H\, {\dot T}^3+2 \l(\frac{V_{_{T}}}{V}\r)
+\Gamma\, {\dot T}\, \l(1+{\dot T}^2\r)\r] \l(\frac{A}{1-\dot T^2}\r)
-\l(\frac{\dot T}{a}\r)\, (\nabla^{2}\, B)
+{\dot T}\, \delta\Gamma=0,\label{eq:deltaT}
\er
where $\delta\Gamma$ denotes the first order perturbation in the 
decay rate. 
Upon using \eq{eq:pece}, we find that the equation describing
the conservation of the perturbed energy density of the fluid 
${\delta \rho_{_{F}}}$ can be written as
\br
\dot {\delta \rho_{_{F}}} 
+ 3\, H\; (1+w_{_{F}})\; {\delta \rho_{_{F}}} 
+\l(\frac{1}{a^2}\r)\nabla^{2} {\psi_{_{F}}}
-\l(\frac{\Gamma\, V\, {\dot T}\, 
(2-\dot T^2)}{\l(1-\dot T^2\r)^{3/2}}\r)\, {\dot {\delta T}}
-\l(\frac{\Gamma V_{_{T}}\, 
{\dot T}^2}{\sqrt{1-\dot T^2}}\r)\, {\delta T}\nn\\
+ \l(\frac{\Gamma\, V\, {\dot T}^2}{\l(1-\dot T^2\r)^{3/2}}\r) A
-\l(\frac{(1+w_{_{F}})\,\rho_{_{F}}}{a}\r) \nabla^2 B 
- \l(\frac{V\, {\dot T}^2}{\sqrt{1-\dot T^2}}\r)\, \delta\Gamma=0. 
\label{eq:rhof}
\er
Also, on using Einstein's equations~(\ref{eq:foee0i}) and~(\ref{eq:foeeii}), we 
obtain the following first order (in time) differential equation for the quantity 
$\psi_{_{F}}$ that describes the spatial velocity of the fluid: 
\beq
\dot \psi_{_{F}}+3\, H\, \psi_{_{F}} 
+ w_{_{F}}\,\delta \rho_{_{F}} 
+(1+w_{_{F}})\, \rho_{_{F}}\, A
+\l(\frac{\Gamma\, V\, {\dot T}}{\sqrt{1-\dot T^2}}\r)\, 
{\delta T}=0.
\label{eq:psif} 
\eeq
We should emphasize here that the above equations take into account 
the two possibilities of $\Gamma$ (i.e. it can either be a constant or 
depend on the tachyon) that we had discussed earlier.

%%%%%%%%%%%%%%%%%%%%%%%%%%%%%%%%%%%%%%%%%%%%%%%%%%%%%%%%%%%%%%%%%%%%%%%%%%%%%%%

\subsection{The coupled perturbation variables and the initial 
conditions}\label{sec:pertic}

The variables describing the perturbations in the tachyon 
and the fluid that we shall eventually evolve numerically 
are~\cite{reviews,dimarco-2007}
\br
{\cal Q}_{_{T}} = \delta T\quad\;\;{\rm and}\quad\;\;
{\cal Q}_{_{F}} 
= \l(\frac{\psi_{_{F}}}{\sqrt{\l(1+w_{_{F}}\r)
\rho_{_{F}}}}\r).\label{eq:calQF}
\er
We have already obtained a second order differential equation 
(in time) describing the evolution of ${\cal Q}_{_{T}}$ [viz. 
\eq{eq:deltaT}]. 
A similar equation for ${\cal Q}_{_{F}}$ can be obtained by 
differentiating \eq{eq:psif} with respect to time, and upon 
using \eq{eq:rhof} that describes the conservation of the 
perturbed energy density of the fluid.
Then, on utilizing the first order Einstein's 
equations~(\ref{eq:foee00}),~(\ref{eq:foee0i}) and~(\ref{eq:foeeij})
to eliminate the scalar metric perturbations $A$ and $B$, we can 
arrive at a set of coupled, second order (in time, again!) 
differential equations for the quantities ${\cal Q}_{_{T}}$ and 
${\cal Q}_{_{F}}$.

For simplicity, let us first discuss the case wherein the decay 
rate $\Gamma$ is a constant.
We shall later indicate as to how certain coefficients in the 
differential equations change when $\Gamma$ is assumed to be 
dependent on the tachyon.
Upon suitably using the background equations~(\ref{eq:fe}),~(\ref{eq:ceF}) 
and~(\ref{eq:emT}), we find that the differential 
equations satisfied by the perturbation variables ${\cal Q}_{_{T}}$
and ${\cal Q}_{_{F}}$ can be written as
\beq
{\ddot {\cal Q}}_{_{\alpha}}
+\sum_{\beta}\, \l({\cal F}_{_{\alpha \beta}}\, 
{\dot {\cal Q}}_{_{\beta}} 
+ {\cal G}_{_{\alpha \beta}}\,
{\cal Q}_{_{\beta}}\r) = 0,\label{eq:ceqtqf}
\eeq
where, as before, $(\alpha,\beta)=(T,F)$. 
The coefficients ${\cal F}_{_{\alpha \beta}}$ and 
${\cal G}_{_{\alpha \beta}}$ appearing in the above equations are 
given by
\br
{\cal F}_{_{TT}} 
&=& \l(3\,H+\Gamma\r) \l(1-3\,{\dot T}^2\r)
- 2\, {\dot T} \l(\frac{V_{_{T}}}{V}\r),\\
%%%%%%%%%%%%%%%%%%%%%%%%%%%%%%%%%
{\cal F}_{_{TF}} 
&=& \l(\frac{4\, \pi\, G}{H}\r)
\l[1-\l(\frac{1-{\dot T}^2}{w_{_{F}}}\r)\r]\, {\dot T}\,
\sqrt{(1+w_{_{F}})\, \rho_{_{F}}}\,,\\
%%%%%%%%%%%%%%%%%%%%%%%%%%%%%%%%%
{\cal F}_{_{FT}} 
&=&\l(\frac{V\,{\dot T}}{\sqrt{1-{\dot T}^2}}\r)\!
\l[\l(\frac{4\, \pi\, G}{H}\r)\!
\l\{1-\l(\frac{w_{_{F}}}{1-{\dot T}^2}\r)\!\r\} \sqrt{(1+w_{_{F}})\, \rho_{_{F}}}
+\l\{1+w_{_{F}}\! \l(\frac{2- {\dot T}^2}{1-{\dot T}^2}\r)\!\r\}\!
\l(\frac{\Gamma}{\sqrt{\l(1+w_{_{F}}\r)\, \rho_{_{F}}}}\r)\r],\\
%%%%%%%%%%%%%%%%%%%%%%%%%%%%%%%%%
{\cal F}_{_{FF}}
&=& 3\, H + \l(\frac{\Gamma\,V\,{\dot T}^2}{\sqrt{1-{\dot T}^2}}\r)
\l(\frac{1}{\rho_{_{F}}}\r),\\
%%%%%%%%%%%%%%%%%%%%%%%%%%%%%%%%%
{\cal G}_{_{TT}} 
&=&\l(1-\dot T^2\r)
\l[\l(\frac{{\rm d}^2{\rm ln}V}{{\rm d}T^2}\r)-\l(\frac{1}{a^2}\r)
\nabla^{2}\r]\nn\\
&+& \l(\frac{4\, \pi\, G}{H}\r) \l(\frac{V\,\dot T}{\sqrt{1-\dot T^2}}\r)
\l[\l(2-{\dot T}^2\r) \l(\frac{2\, V_{_{T}}}{V}\r)
+ \l(2+{\dot T}^2\r) \l(3\,H+\Gamma\r)\, {\dot T}
-\l(\frac{\Gamma\, {\dot T}}{w_{_{F}}}\r) \l(1-{\dot T}^2\r)\r]\nn\\ 
&-& \l(\frac{4\, \pi\, G}{H}\r)^2 \l[\l(\frac{2\, V^2\, {\dot T}^4}{1-{\dot T}^2}\r)
- \l(\frac{V\,{\dot T}^2}{\sqrt{1-{\dot T}^2}}\r)
\l\{1+\l(\frac{1-{\dot T}^2}{w_{_{F}}}\r)\r\}
\l(1+w_{_{F}}\r)\,\rho_{_{F}}\r],\\
%%%%%%%%%%%%%%%%%%%%%%%%%%%%%%%%%%
{\cal G}_{_{TF}} 
&=&\l(\frac{4\, \pi\, G}{H}\r)
\Biggl[- \l(\frac{3\, H\, {\dot T}}{2}\r)
\l\{1+\l(\frac{1-\dot T^2}{w_{_{F}}}\r)\r\} \l(1+w_{_{F}}\r)
+\l(\frac{1}{2\,\rho_{_{F}}}\r) \l\{1-\l(\frac{1-{\dot T}^2)}{w_{_{F}}}\r)\r\}
\l(\frac{\Gamma\, V\, {\dot T}^3}{\sqrt{1-{\dot T}^2}}\r)\Biggr. \nn\\
&-& \Biggl. \l\{6\, H\,{\dot T}^3 + 2\,\l(\frac{V_{_{T}}}{V}\r)
+(\Gamma\,{\dot T}) \l(1+{\dot T}^2\r)\r\}\Biggr] \sqrt{\l(1+w_{_{F}}\r)\,\rho_{_{F}}}\nn\\ 
&+&\l(\frac{4\, \pi\, G}{H}\r)^2
\l[\l(\!\frac{2\, V\,{\dot T}^3}{\sqrt{1-{\dot T}^2}}\!\r)
+ \l\{1+\l(\!\frac{1-{\dot T}^2}{w_{_{F}}}\!\r)\r\} {\dot T}\,
\l[(1+w_{_{F}})\,\rho_{_{F}}\r]\r]
\sqrt{\l(1+w_{_{F}}\r)\, \rho_{_{F}}}\,,\\
%%%%%%%%%%%%%%%%%%%%%%%%%%%%%%%%%%
{\cal G}_{_{FT}} 
&=& -\l(\frac{4\, \pi\, G}{H}\r)\, 
\Biggl[\l\{1+\l(\frac{w_{_{F}}}{1-{\dot T}^2}\r)\r\}
\l[\l(1+w_{_{F}}\r)\,\rho_{_{F}}\r]\, V_{_{T}} \sqrt{1-{\dot T}^2}\Biggr. \nn\\
&+& \Biggl.\l(\!\frac{3\, H\,V\, {\dot T}}{\sqrt{1-{\dot T}^2}}\!\r)\,
\l(1+w_{_{F}}\r)^2 \rho_{_{F}}
- \l(\!\frac{\Gamma\, V^2\, {\dot T}^3}{1-{\dot T}^2}\r)
\l\{1-\l(\!\frac{w_{_{F}}\, {\dot T}^2}{1-{\dot T}^2}\r)\r\}\Biggr] 
\l(\frac{1}{\sqrt{\l(1+w_{_{F}}\r)\, \rho_{_{F}}}}\r)\nn\\ 
&-&\l(\frac{4\, \pi\, G}{H}\r)^2
\Biggl[\l(\frac{V^2\, {\dot T}^3}{1-{\dot T}^2}\r)
\l\{1+\l(\frac{w_{_{F}}}{1-{\dot T}^2}\r)\r\}
\l[\l(1+w_{_{F}}\r)\,\rho_{_{F}}\r]
-\l(\frac{2\, V\, {\dot T}}{\sqrt{1-{\dot T}^2}}\r)
\l[\l(1+w_{_{F}}\r)\, \rho_{_{F}}\r]^{2}\Biggr]
\l(\frac{1}{\sqrt{\l(1+w_{_{F}}\r)\, \rho_{_{F}}}}\r)\nn\\
&+& \l(\frac{\Gamma\, V}{\sqrt{1-\dot T^2}}\r)
\l[\l(\frac{V_{_{T}}}{V}\r)\, \l(1-{\dot T}^2\r)
- \l(\frac{V_{_{T}}}{V}\r)\, w_{_{F}}\, {\dot T}^2
+\Gamma\, {\dot T} -3\, H\, w_{_{F}}\, {\dot T}\r]
\l(\frac{1}{\sqrt{\l(1+w_{_{F}}\r)\, \rho_{_{F}}}}\r),\\
%%%%%%%%%%%%%%%%%%%%%%%%%%%%%%%%%%%
{\cal G}_{_{FF}} 
&=& -\l(\frac{w_{_{F}}}{a^2}\r) \nabla^{2} 
+ \l(\frac{9\, H^2}{4}\r) \l(1-w_{_{F}}^2\r)\nn\\ 
&+& \l(\!\frac{4\, \pi\, G}{H}\r)
\Biggl[\l(\!\frac{3\, H}{2}\r) \l(1+w_{_{F}}\r) \l(1+3\,w_{_{F}}\r)  \rho_{_{F}}
+\l(\frac{3\, H}{2}\r)\! \l(\frac{V\, {\dot T}^2}{\sqrt{1-{\dot T}^2}}\r)\l(w_{_{F}}-1\r)
-\l(\frac{\Gamma\, V\, {\dot T}^2}{\sqrt{1-{\dot T}^2}}\r)
\l\{1-\l(\frac{w_{_{F}} {\dot T}^2}{1-\dot T^2}\r)\r\}\Biggr]\nn\\
&-&\l(\frac{4\, \pi\, G}{H}\r)^{2}
\Biggl[\l(\frac{V\, {\dot T}^2}{\sqrt{1-{\dot T}^2}}\r)
\l\{1+\l(\frac{w_{_{F}}}{1-\dot T^2}\r)\r\} \l[\l(1+w_{_{F}}\r)\, \rho_{_{F}}\r]
+ 2 \l[\l(1+w_{_{F}}\r)\, \rho_{_{F}}\r]^2\Biggr]\nn\\
&-& \l(\frac{\Gamma\, V\, {\dot T}}{\sqrt{1-\dot T^2}}\r)
\Biggl[\l(\frac{\Gamma\, {\dot T}}{2\, \rho_{_{F}}}\r) \l(2-{\dot T}^{2}\r)
+\l(\frac{V_{_{T}}}{V}\r) \l(\frac{1-{\dot T}^{2}}{\rho_{_{F}}}\r)
- \l(\frac{3\, H\, {\dot T}}{2\, \rho_{_{F}}}\r) \l(w_{_{F}}+{\dot T}^{2}\r)
+ \l(\frac{\Gamma\, V\, {\dot T}^3}{4\, \rho_{_{F}}^2\, \sqrt{1-\dot T^2}}\r)\Biggr].
\er
%%%%%%%%%%%%%%%%%%%%%%%%%%%%%%%%%%%%
As we had pointed out, the coefficients ${\cal F}_{_{\alpha\beta}}$ 
and ${\cal G}_{_{\alpha\beta}}$ above have been derived assuming 
that the decay rate $\Gamma$ is a constant. 
For the case wherein the decay rate is a function of the tachyon, 
the coefficients ${\cal G}_{_{TT}}$ and ${\cal G}_{_{FT}}$ are 
modified to 
\br
{\cal G}_{_{TT}} 
& \to & {\cal G}_{_{TT}}
+ \l(1-{\dot T}^2\r)\, {\dot T}\; \Gamma_{_{T}},\\
{\cal G}_{_{FT}} 
& \to & {\cal G}_{_{FT}}
+ \l(\frac{V\,{\dot T}^2}{\sqrt{1-{\dot T}^2}}\r)\, 
\l(\frac{w_{_{F}}}{\sqrt{\l(1+w_{_{F}}\r)\, \rho_{_{F}}}}\r)\, 
\Gamma_{_{T}},
\er
where $\Gamma_{_{T}}\equiv \l({\rm d}\,\Gamma/{\rm d}T\r)$. 
The other coefficients remain unaffected in this case.

We should mention here that we have checked that we indeed
recover the standard equations in the different limiting cases 
from the coupled equations~(\ref{eq:ceqtqf}) for ${\cal Q}_{_{T}}$ 
and ${\cal Q}_{_{F}}$.
For instance, when the fluid is ignored and the decay rate is 
assumed to vanish, the equation for the variable ${\cal Q}_{_{T}}$ 
reduces to the equation that has been derived earlier in the
literature for describing the perturbation in the tachyon (see, for 
instance, Refs.~\cite{rajeev-2007,steer-2004,wti}).
Moreover, when the tachyon is assumed to vanish and its coupling 
to the fluid is also ignored, we find that we arrive at the well-known 
equation governing the fluid perturbation, as required~\cite{texts,reviews}.

%%%%%%%%%%%%%%%%%%%%%%%%%%%%%%%%%%%%%%%%%%%%%%%%%%%%%%%%%%%%%%%%%%%%%%%%%%%%%%%

\section{Effects of reheating on the scalar power 
spectrum}\label{sec:ep}

In this section, we shall discuss the effects of reheating on the
large scale curvature perturbations and the scalar power spectrum.
We shall also investigate the behavior of the non-adiabatic pressure
perturbations at large scales, when reheating is achieved through 
the two different choices for the decay rate.
 
%%%%%%%%%%%%%%%%%%%%%%%%%%%%%%%%%%%%%%%%%%%%%%%%%%%%%%%%%%%%%%%%%%%%%%%%%%%%%%%

\subsection{Evolution of the curvature perturbations and the 
scalar power spectrum}\label{sec:ecp}

We shall solve the coupled equations~(\ref{eq:ceqtqf}) for the two 
variables ${\cal Q}_{_{T}}$ and ${\cal Q}_{_{F}}$ numerically.
Imposing the standard Bunch-Davies initial conditions, we shall evolve 
the perturbations from the sub-Hubble to the super-Hubble scales, and 
analyse the effects of the transition from inflation to the radiation 
dominated epoch on the evolution of the large scale perturbations.
We shall compute the scalar power spectrum at the end of inflation
and soon after reheating is complete.
Let us begin by quickly summarizing the essential quantities of 
interest and the initial conditions that we shall be imposing.

Let ${\cal R}_{_{T}}$ and ${\cal R}_{_{F}}$ denote the curvature 
perturbations associated with the tachyon and the fluid, 
respectively.
We find that these curvature perturbations are related to 
the variables ${\cal Q}_{_{T}}$ and ${\cal Q}_{_{F}}$ as 
follows~\cite{reviews,rajeev-2007,steer-2004,wti}:
\br
{\cal R}_{_{T}}
&=& \l(\frac{H}{\dot T}\r)\, \delta T
\ =\ \l(\frac{H}{\dot T}\r)\, {\cal Q}_{_{T}},\\
{\cal R}_{_{F}} 
&=&-\l(\frac{H}{\rho_{_{F}}+p_{_{F}}}\r)\, \psi_{_{F}}
\ =\ -\l(\frac{H}{\sqrt{\l(1+w_{_{F}}\r)\,\rho_{_{F}}}}\r)\, 
{\cal Q}_{_{F}}.
\er
The total curvature perturbation of the system, say, ${\cal R}$, 
can then be expressed as a weighted sum of the individual curvature 
perturbations in the following fashion (see, for instance, 
Refs.~\cite{malik-2001,malik-2005}):
\beq
{\cal R}=\sum\limits_{\alpha} 
\l(\frac{\rho_{_{\alpha}}+p_{_{\alpha}}}{\rho+p}\r)\, 
{\cal R}_{_{\alpha}}.
\eeq
Therefore, for our system of interest, the total curvature 
perturbation is given by
\beq
{\cal R}= H\, \l[\l(\frac{V\, {\dot T}^{2}}{\sqrt{1-{\dot T}^2}}\r)
+\l[\l(1+w_{_{F}}\r)\, \rho_{_{F}}\r]\r]^{-1}
\l[\l(\frac{V\, {\dot T}}{\sqrt{1-{\dot T}^2}}\r)\, {\cal Q}_{_{T}} 
-\sqrt{\l[\l(1+w_{_{F}}\r)\, \rho_{_{F}}\r]}\; {\cal Q}_{_{F}}\r].
\eeq
The scalar power spectrum is then defined in terms of the Fourier
mode ${\cal R}_{k}$ of the total curvature perturbation as
\beq
{\cal P}_{_{\rm S}}(k)=\l(\frac{k^{3}}{2\, \pi^{2}}\r)\,
\vert{\cal R}_{\rm k}\vert^{2}.
\eeq

The Mukhanov-Sasaki variables, say, $v_{_{T}}$ and $v_{_{F}}$, 
associated with the tachyon and the fluid are related to the 
corresponding curvature perturbations as follows:
\beq
v_{_{T}}=\l({\cal R}_{_{T}}\, z_{_{T}}\r)
\qquad{\rm and}\qquad
v_{_{F}}=\l({\cal R}_{_{F}}\, z_{_{F}}\r),
\eeq
where the quantities $z_{_{T}}$ and $z_{_{F}}$ are given 
by~\cite{texts,reviews,rajeev-2007,steer-2004}
\beq
z_{_{T}}=\l(\frac{3}{8\,\pi\,G}\r)^{1/2}
\l(\frac{a\, {\dot T}}{\sqrt{1-{\dot T}^2}}\r)
\qquad{\rm and}\qquad
z_{_{F}}=\l(\frac{a}{H}\r) \l[\frac{\l(1+w_{_{F}}\r)\, 
\rho_{_{F}}}{w_{_{F}}}\r]^{1/2}.
\eeq
As is usually done, we shall impose the following initial conditions 
on the Fourier modes of the perturbation variables $v_{_{T}}$ and 
$v_{_{F}}$:
\beq
v_{k}=\l(\frac{1}{2\, \omega_{k}}\r)^{1/2}
\qquad{\rm and}\qquad
{\dot v}_{k}=-\l(\frac{i}{a}\r)\, \l(\frac{\omega_{k}}{2}\r)^{1/2},
\eeq
where 
\beq
\omega_{k}^{2}=\l[\l(k\, c_{_{\rm S}}\r)^{2} 
-a^{2} \l\{\l({\ddot z}/z\r)+\l(H\, {\dot z}/{z}\r)\r\}\r],
\eeq
with $c_{_{\rm S}}^2=(1-{\dot T}^2)$ in the case of the tachyon and
$c_{_{\rm S}}^2=w_{_{F}}$ in the case of the fluid.
Also, we shall impose these conditions when the modes are well inside 
the Hubble radius during the inflationary epoch.
We shall transform these initial conditions to the corresponding 
conditions on the variables ${\cal Q}_{_{T}}$ and ${\cal Q}_{_{F}}$
when numerically integrating the coupled equations~(\ref{eq:ceqtqf}).

\begin{figure}[!htb]
\begin{center}
\vskip 15pt\hskip 15pt
\resizebox{200pt}{130pt}{\includegraphics{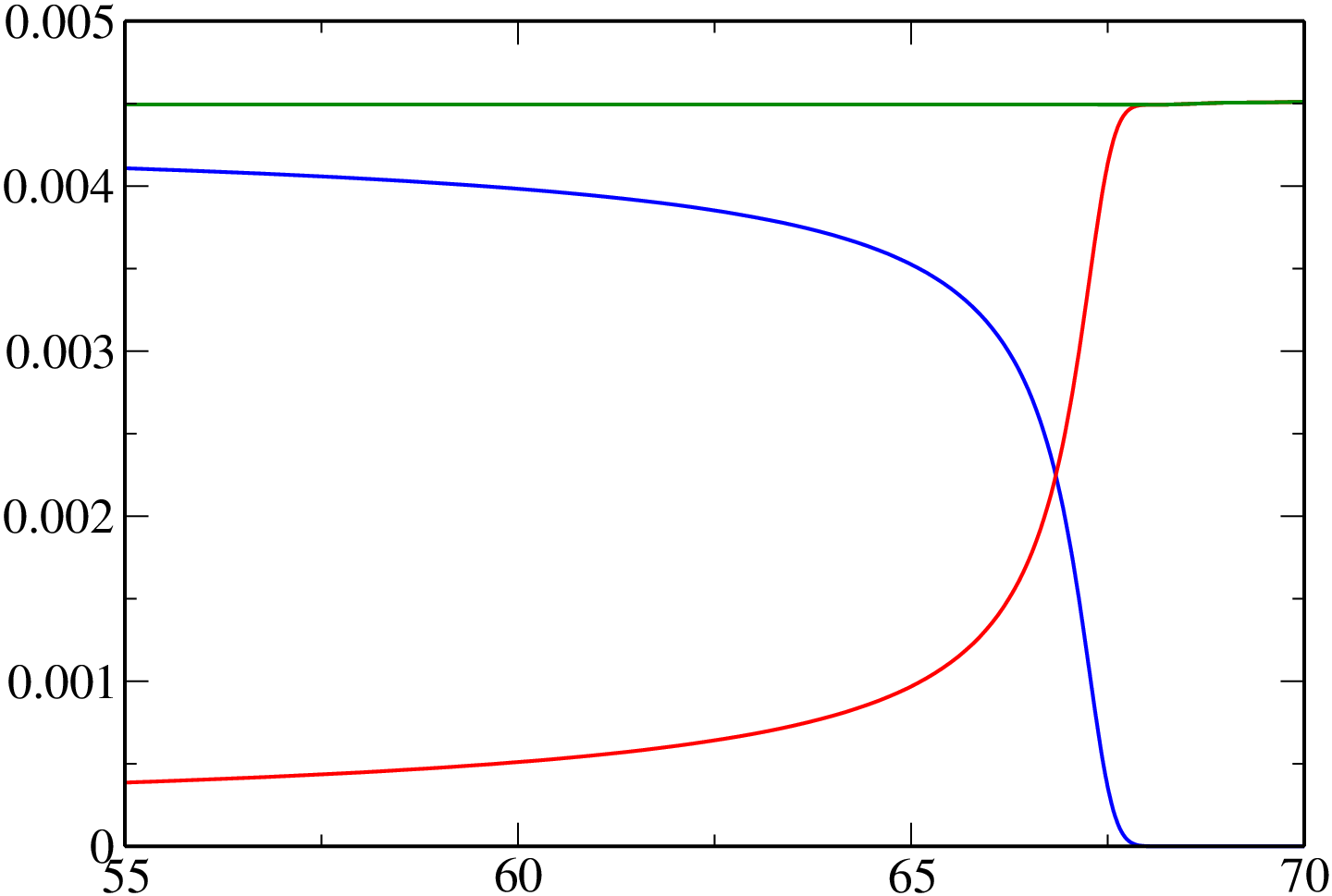}}
\hskip 25pt
\resizebox{200pt}{130pt}{\includegraphics{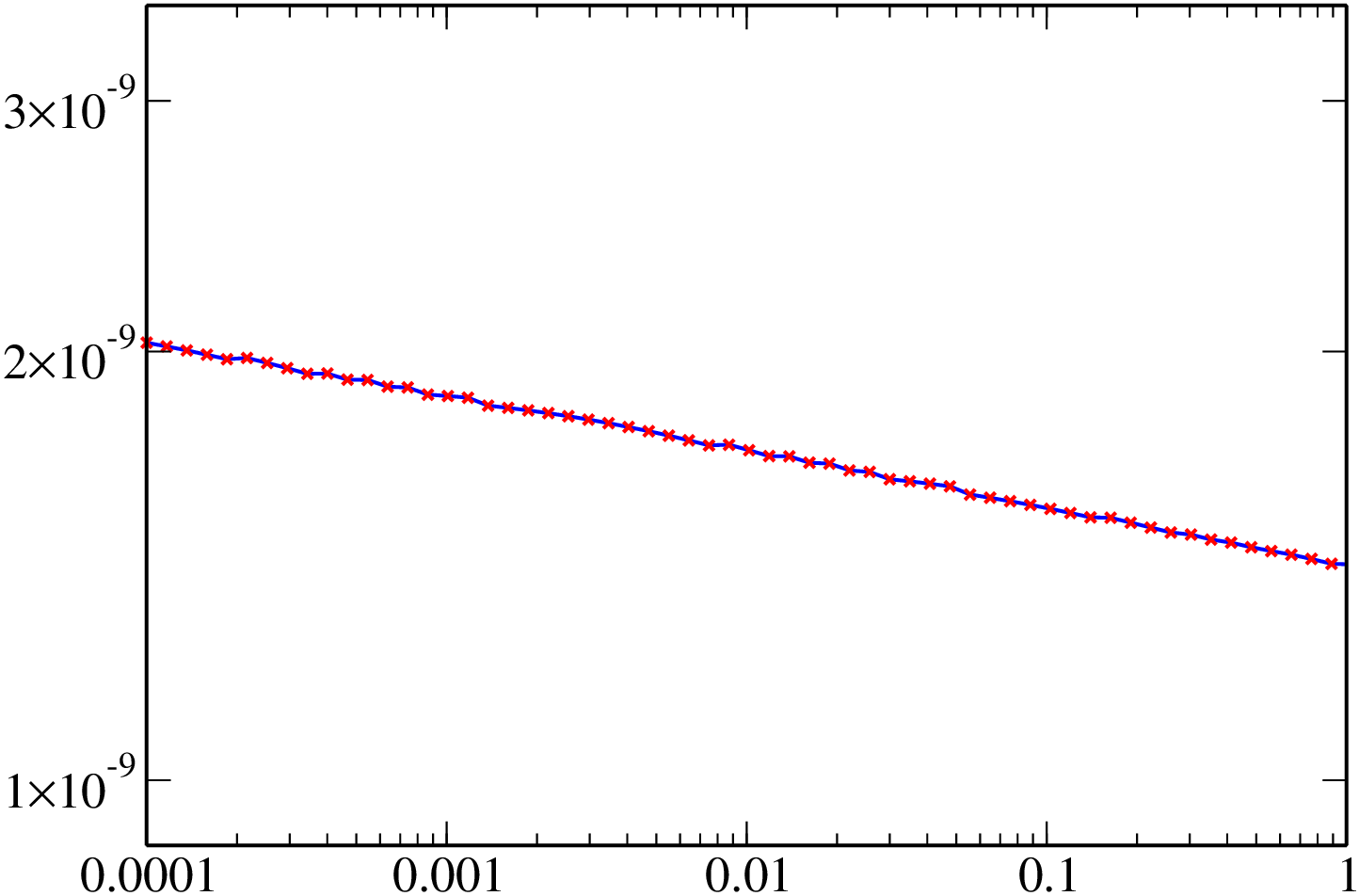}}
\vskip -103 true pt 
\hskip -245 true pt $\vert{\cal R}_{_{T}}\vert, 
\vert{\cal R}_{_{\gamma}}\vert, \vert{\cal R}\vert$
\hskip 210 true pt ${\cal P}_{_{\rm S}}(k)$
\vskip 108 pt\hskip 15pt
\resizebox{200pt}{130pt}{\includegraphics{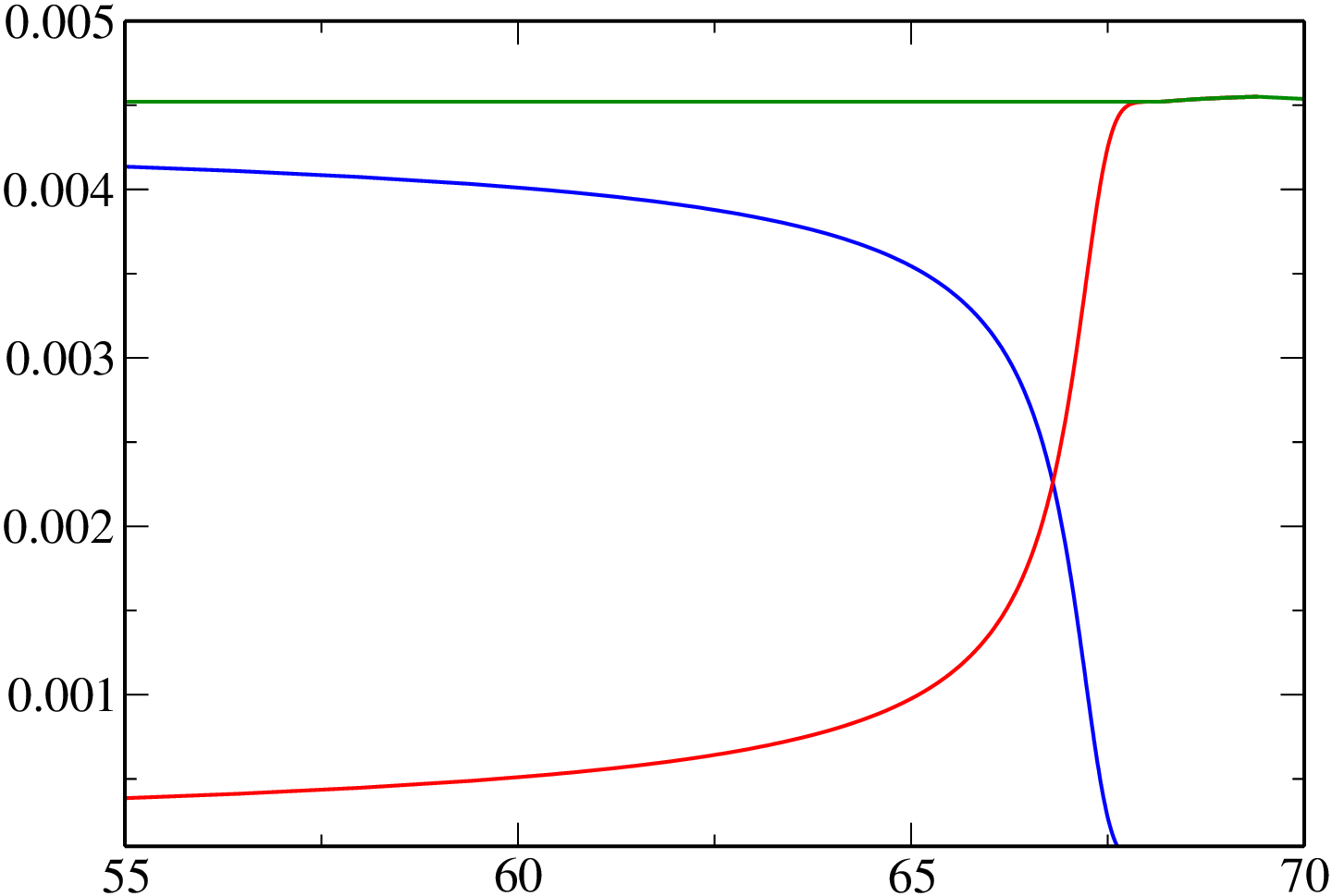}}
\hskip 25pt
\resizebox{200pt}{130pt}{\includegraphics{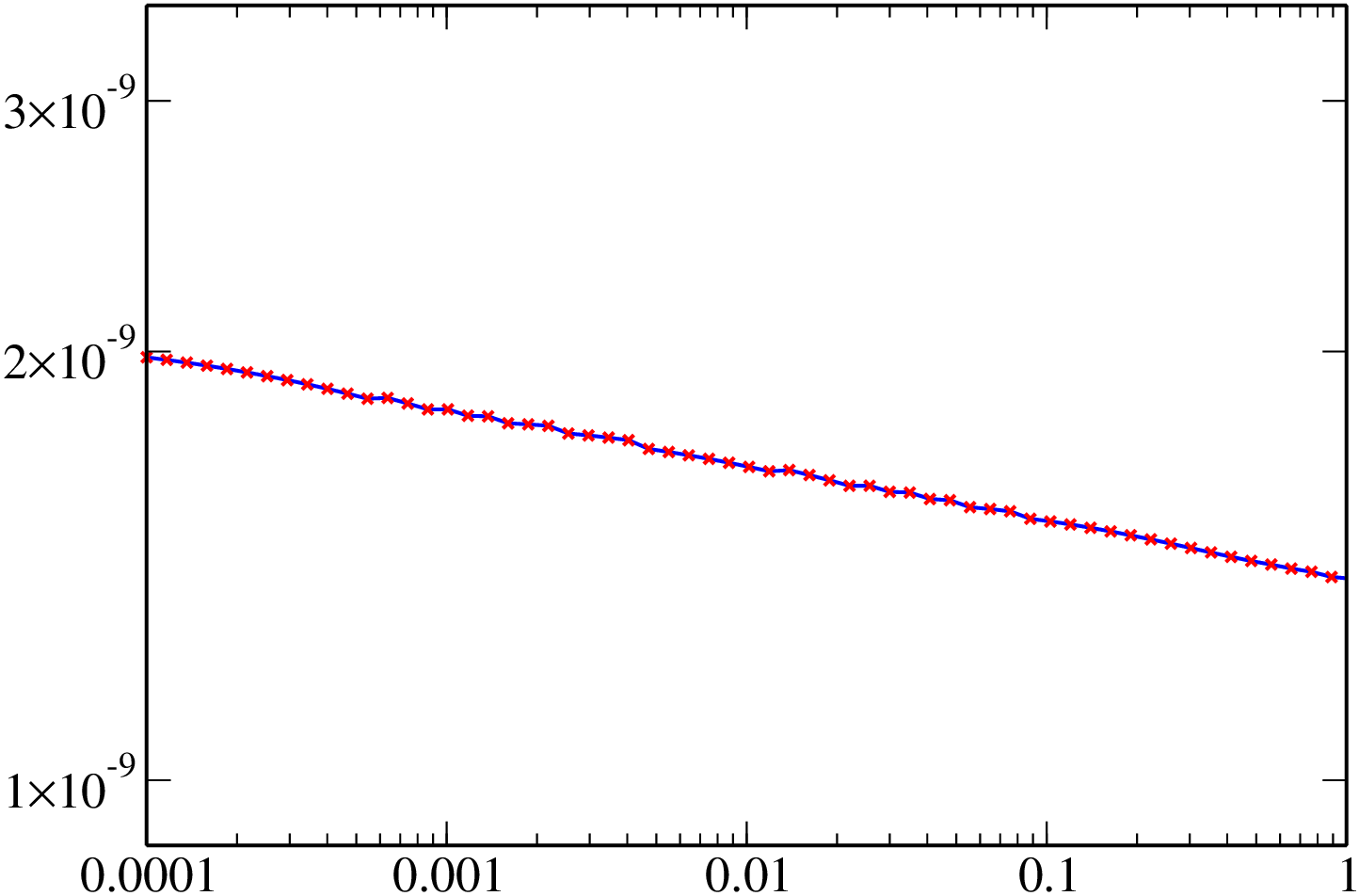}}
\vskip -102 true pt 
\hskip -245 true pt $\vert{\cal R}_{_{T}}\vert, 
\vert{\cal R}_{_{\gamma}}\vert, \vert{\cal R}\vert$
\hskip 210 true pt ${\cal P}_{_{\rm S}}(k)$
\vskip 95 pt
\hskip 25 true pt $N$ \hskip 215 true pt $k$
\vskip 5pt
\caption{\label{fig:modepsdbif}In the left column, the evolution 
of the amplitudes of the curvature perturbations associated with 
the tachyon ($\vert{\cal R}_{_{T}}\vert$, in blue) and radiation 
($\vert{\cal R}_{_{\gamma}}\vert$, in red) as well as the weighted 
sum of the two ($\vert{\cal R}\vert$, in green) has been plotted 
as a function of the number of $e$-folds~$N$ for the two possible 
cases of $\Gamma$, viz. $\Gamma_{1}={\rm constant}$ (on top) and 
$\Gamma_{2}=\Gamma(T)$ (at the bottom).  
In these plots, we have chosen a typical cosmological scale with 
the wavenumber $k=0.1\;{\rm Mpc^{-1}}$.
Clearly, the amplitude of the total curvature perturbation remains 
constant across the transition in both the cases.
In the right column, the scalar power spectrum has been plotted 
towards the end of inflation (in blue) and soon after complete 
reheating has been achieved (in red) for the two cases of $\Gamma$, 
as mentioned above.
These plots unambiguously illustrate that the amplitude of the nearly 
scale invariant power spectrum is unaffected in both the cases.
(In these plots, the blue line lies right beneath the red one!)
Note that all these plots correspond to the potential $V_{1}(T)$ with
values for the various parameters and the initial conditions that are 
listed in Tables~\ref{tab:beiaf} and~\ref{tab:beipf}.
We find that a very similar behavior occurs for the potential $V_{2}(T)$.}
\end{center}
\end{figure}

We shall focus on modes that correspond to cosmological scales
today.
We shall choose our parameters in such a fashion that modes spread
over four orders of magnitude (say, $10^{-4}\lesssim k\lesssim 1\;
{\rm Mpc}^{-1}$) leave the Hubble radius during the early stages of 
inflation (they actually leave during $5\lesssim N \lesssim 15$).
As we have discussed earlier, for our choice of the parameters and 
initial conditions on the background variables, reheating is achieved 
after about $65$ odd $e$-folds (also see Fig.~\ref{fig:eewdbif}).
Moreover, the transition from inflation to radiation domination occurs
within about $2$-$3$ $e$-folds after inflation has terminated.
We evaluate the scalar power spectrum towards the end of inflation
as well as soon after the transition to the radiation dominated
epoch is complete.
In the left column of Fig.~\ref{fig:modepsdbif}, we have plotted the 
evolution of the amplitudes of the individual (viz. 
$\vert{\cal R}_{_{T}}\vert$ and $\vert{\cal R}_{_{\gamma}}\vert$ 
corresponding to the tachyon and radiation, respectively) as well 
as the total curvature perturbation (i.e. $\vert{\cal R}\vert$) 
for a typical cosmological scale (we have chosen $k=0.1\;{\rm Mpc}^{-1}$) 
across the transition from inflation to radiation domination.
And, in the right column of the figure, we have plotted the scalar
power spectrum (constructed out of the total curvature perturbation)
evaluated before and after the transition.
It is evident from the plots that the amplitude of the total 
curvature perturbation remains unaffected during reheating when 
the decay rate $\Gamma$ is either a constant or is a function of 
the tachyon. 
These results confirm similar conclusions that have been arrived
at earlier based on the first order (in time) super-Hubble 
equations~\cite{matarrese-2003}.

%%%%%%%%%%%%%%%%%%%%%%%%%%%%%%%%%%%%%%%%%%%%%%%%%%%%%%%%%%%%%%%%%%%%%%%%%%%%%%%

\subsection{Evolution of the entropy perturbations}\label{sec:eep}

In this section, we shall discuss the evolution of the entropic (i.e.
the non-adiabatic pressure) perturbations across the transition from
inflation to the radiation dominated epoch. 

Let us recall the key quantities and equations.
In the UCG, the gauge invariant Bardeen potential~$\Phi$ is given by
\beq
\Phi=A+a\,\l({\dot B}+H\, B\r).
\eeq
Upon using the first order Einstein's equations~(\ref{eq:foee00}),~(\ref{eq:foeeii}) 
and~(\ref{eq:foeeij}), it can be shown that the Bardeen potential satisfies the 
differential equation~\cite{texts,reviews}
\beq
{\ddot \Phi} 
+ \l(4+3\, c_{_{\rm A}}^2\r)\, H\, {\dot \Phi}
+\l[2\, {\dot H}+3\, H^2\, \l(1+ c_{_{\rm A}}^2\r)\r]\, \Phi
-\l(\frac{c_{_{\rm A}}^2}{a^{2}}\r)\, \nabla^{2}\,\Phi 
= \l(4\,\pi\, G\r)\, \delta p^{_{\rm NA}},
\label{eq:emPhi}
\eeq
where $c_{_{\rm A}}^{2}=\l({\dot p}/{\dot \rho}\r)$ denotes the 
adiabatic speed of sound in the composite system, and the quantity 
$\delta p^{_{\rm NA}}$ is the non-adiabatic pressure perturbation 
defined as
\beq
\delta p^{_{\rm NA}}
=\l(\delta p-c_{_{\rm A}}^2\, \delta \rho\r).
\eeq
Using Einstein's equations~(\ref{eq:foee0i}) and~(\ref{eq:foeeij}), 
the total curvature perturbation ${\cal R}$ of the system can be 
written in terms of the Bardeen potential~$\Phi$ as 
follows~\cite{texts,reviews}:
\beq
{\cal R}=\Phi + \l(\frac{2\, \rho}{3\, H}\r)
\l(\frac{{\dot \Phi}+H\, \Phi}{\rho+p}\r).
\label{eq:R}
\eeq
On substituting this expression for the curvature perturbation in 
\eq{eq:emPhi} that governs the evolution of the 
Bardeen potential~$\Phi$ and making use of the background equations, 
one can arrive at the well-known equation
\beq
{\dot {\cal R}}=-\l(\frac{H}{\dot H}\r)
\l[\l(4\, \pi\, G\r)\, \delta p^{_{\rm NA}} 
+\l(\frac{c_{_{\rm A}}^{2}}{a^{2}}\r)\, \nabla^{2}\Phi\r].
\label{eq:emRdot} 
\eeq
This equation clearly suggests that, at super-Hubble scales, any
change in the amplitude of the curvature perturbations has to be 
due to the non-trivial evolution of the non-adiabatic pressure 
perturbation $\delta p^{_{\rm NA}}$.

The perfect fluid, by definition, does not possess any intrinsic 
entropy perturbation. 
Therefore, the non-adiabatic pressure perturbation $\delta p^{_{\rm NA}}$ 
of our composite system depends on the intrinsic non-adiabatic pressure 
perturbation of the tachyon, say, $\delta p_{_{T}}^{_{\rm NA}}$, and 
the relative non-adiabatic pressure perturbation between the tachyon and 
the perfect fluid (i.e. the isocurvature perturbation), which we shall 
denote as $\delta p_{_{TF}}^{_{\rm NA}}$.
Hence, we can write 
\beq
\delta p^{_{\rm NA}} 
= \l(\delta p_{_{T}}^{_{\rm NA}}+\delta p_{_{TF}}^{_{\rm NA}}\r),
\eeq
where the quantity $\delta p_{_{T}}^{_{\rm NA}}$ is defined as
(see, for instance, Ref.~\cite{rajeev-2007}) 
\beq
\delta p_{_{T}}^{_{\rm NA}}
=\l(\delta p_{_{T}}-c_{_{\rm T}}^2\, \delta \rho_{_{T}}\r)
\eeq
with $c_{_{\rm T}}^2=\l({\dot p}_{_{T}}/{\dot \rho}_{_{T}}\r)$.
The relative non-adiabatic pressure perturbation 
$\delta p_{_{TF}}^{_{\rm NA}}$ is then given by (see, for instance, 
Refs.~\cite{matarrese-2003,gordon-2001,bartolo-2004})
\beq
\delta p_{_{TF}}^{_{\rm NA}}
=\l(3\,H+\Gamma\r) \l(w_{_{F}}-c_{_{\rm T}}^2\r) \l(\frac{\rho_{_{T}}+p_{_{T}}}{\rho+p}\r)
\Bigl[\l(\rho_{_{F}}+p_{_{F}}\r)-\l(\Gamma/3\, H\r) \l(\rho_{_{T}}+p_{_{T}}\r)\Bigr]
\l[\l(\frac{\delta \rho_{_{T}}}{\dot \rho_{_{T}}}\r)
-\l(\frac{\delta \rho_{_{F}}}{\dot \rho_{_{F}}}\r)\r].
\eeq
Upon using the expressions~(\ref{eq:deltarho}) and~(\ref{eq:deltap}), 
and Einstein's equations~(\ref{eq:foee00}),~(\ref{eq:foeeii}) 
and~(\ref{eq:foeeij}), we find that we can write the quantities 
$\delta \rho_{_{T}}$, $\delta p_{_{T}}$ and $\delta \rho_{_{F}}$ in 
terms of the variables ${\cal Q}_{_{T}}$ and ${\cal Q}_{_{F}}$ as 
follows:
\br
\delta \rho_{_{T}} 
&=& \l[\l(\frac{V_{_{T}}}{\sqrt{1-{\dot T}^2}}\r)
-\l(\frac{4\,\pi\, G}{H}\r)\, 
\l(\frac{V^{2}\, {\dot T}^{3}}{\l(1-{\dot T}^2\r)^{2}}\r)\,\r] {\cal Q}_{_{T}}
+\l(\frac{V\, {\dot T}}{\l(1-{\dot T}^{2}\r)^{3/2}}\r) {\dot {\cal Q}}_{_{T}}\nn\\
&+&\,\l(\frac{4\,\pi\, G}{H}\r)\, 
\l(\frac{V\, {\dot T}^{2}}{\l(1-{\dot T}^{2}\r)^{3/2}}\r)
\sqrt{\l(1+w_{_{F}}\r)\,\rho_{_{F}}}\;\, {\cal Q}_{_{F}},\\
\delta p_{_{T}}
&=& -\l[V_{_{T}}\, \sqrt{1-{\dot T}^2}+\l(\frac{4\,\pi\, G}{H}\r)\, 
\l(\frac{V^{2}\, {\dot T}^{3}}{1-{\dot T}^2}\r)\r] {\cal Q}_{_{T}}
+\l(\frac{V\, {\dot T}}{\sqrt{1-{\dot T}^{2}}}\r) {\dot {\cal Q}}_{_{T}}\nn\\
&+& \l(\frac{4\,\pi\, G}{H}\r)\, 
\l(\frac{V\, {\dot T}^{2}}{\sqrt{1-{\dot T}^2}}\r)\, 
\sqrt{\l(1+w_{_{F}}\r)\,\rho_{_{F}}}\;\, {\cal Q}_{_{F}},\\
\delta \rho_{_{F}} 
&=& -\l[\l(\frac{4\,\pi\, G}{H}\r)\,
\l(\frac{V\, {\dot T}}{\sqrt{1-{\dot T}^{2}}}\r)
\l(1+w_{_{F}}\r)\, \l(\frac{\rho_{_{F}}}{w_{_{F}}}\r) 
+ \l(\frac{\Gamma\, V\, {\dot T}}{w_{_{F}}\, 
\sqrt{1-{\dot T}^{2}}}\r)\r]\, {\cal Q}_{_{T}}\nn\\ 
&-& \l[\l(\frac{3\, H}{2\, w_{_{F}}}\r) \l(1-w_{_{F}}\r)
\sqrt{\l(1+w_{_{F}}\r)\, \rho_{_{F}}}
+\l(\frac{\Gamma\, V\, {\dot T}^{2}}{\sqrt{1-{\dot T}^2}}\r)
\l(\frac{1+w_{_{F}}}{4\, w_{_{F}}^{2}\,\rho_{_{F}}}\r)^{1/2}
- \l(\!\frac{4\,\pi\, G}{H}\!\r) \l(\frac{1}{w_{_{F}}}\r)
\l[\l(1+w_{_{F}}\r)\, \rho_{_{F}}\r]^{3/2}\r] {\cal Q}_{_{F}} \nn\\
&-&\l(\frac{1}{w_{_{F}}}\r)\,\sqrt{\l(1+w_{_{F}}\r)\, \rho_{_{F}}}\, 
{\dot {\cal Q}}_{_{F}}.
\er
These three relations allow us to express the non-adiabatic 
pressure perturbations $\delta p_{_{T}}^{_{\rm NA}}$ and
$\delta p_{_{TF}}^{_{\rm NA}}$ in terms of ${\cal Q}_{_{T}}$, 
${\cal Q}_{_{F}}$, and their time derivatives.

\begin{figure}[!htb]
\begin{center}
\vskip 15pt
\resizebox{220pt}{140pt}{\includegraphics{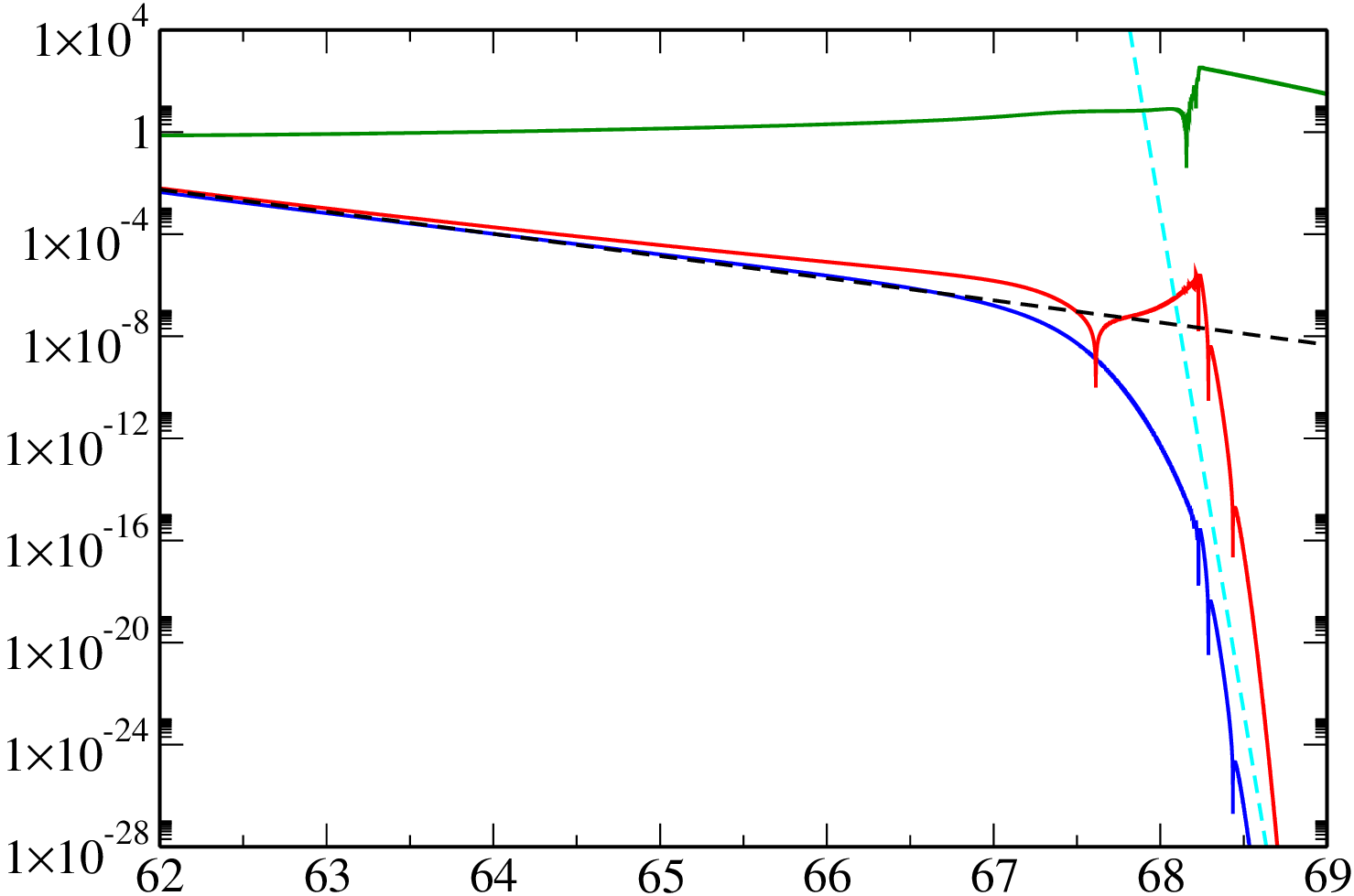}}
\vskip -125 true pt 
\hskip -300 true pt $\vert\Phi\vert, 
\vert\delta p^{_{\rm NA}}_{_{T}}\vert, 
\vert\delta p^{_{\rm NA}}_{_{T\gamma}}\vert$
\vskip 135 pt
\resizebox{220pt}{140pt}{\includegraphics{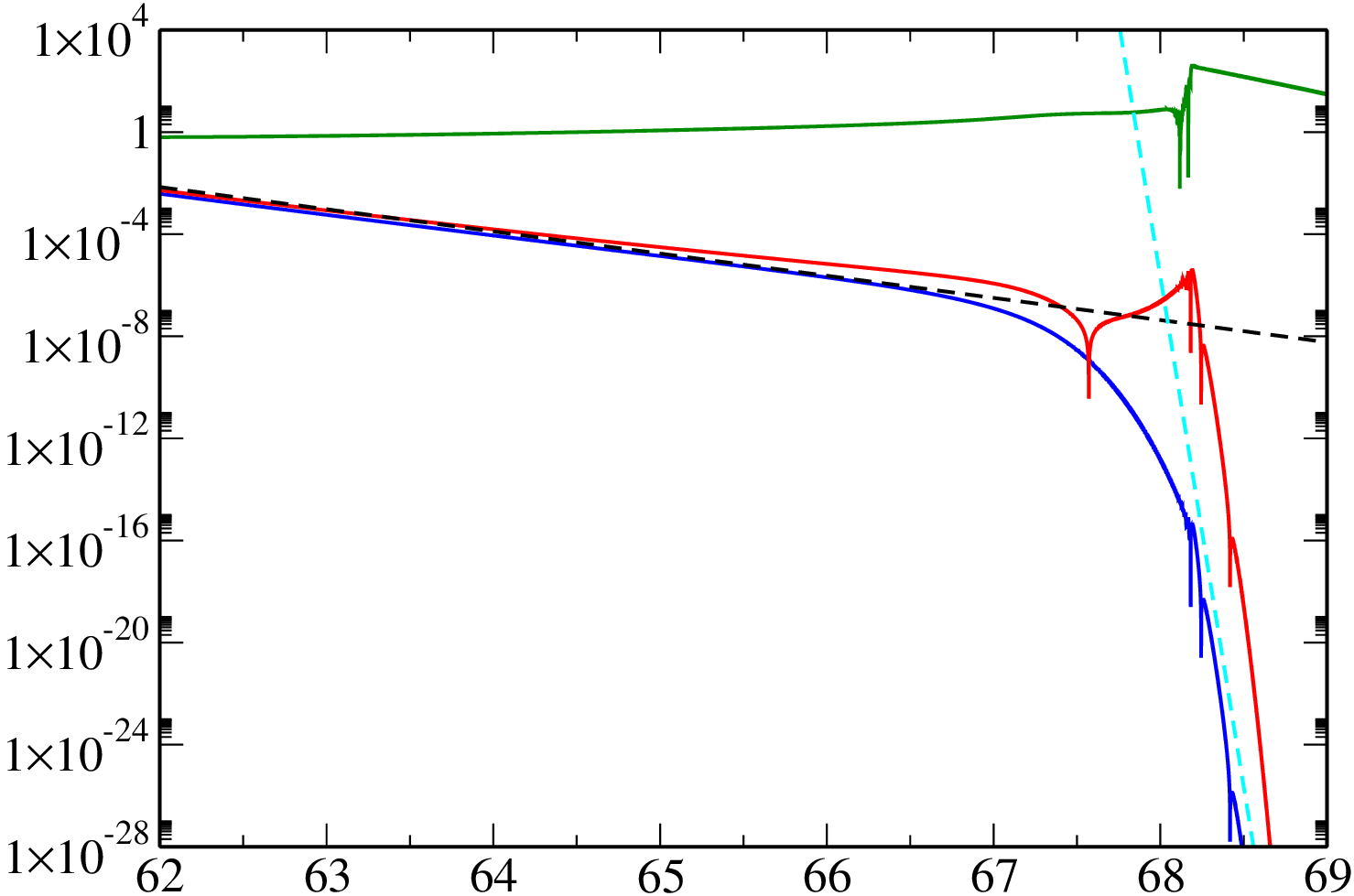}}
\vskip -125 true pt 
\hskip -300 true pt $\vert\Phi\vert, 
\vert\delta p^{_{\rm NA}}_{_{T}}\vert, 
\vert\delta p^{_{\rm NA}}_{_{T\gamma}}\vert$
\vskip 115 pt
\hskip 15 true pt $N$ 
\vskip 5pt
\caption{\label{fig:enapp} The evolution of the amplitudes of the 
Bardeen potential $\vert\Phi\vert$ (in green), the intrinsic entropy
perturbation $\vert\delta p^{_{\rm NA}}_{_{T}}\vert$ (in blue) and the
isocurvature perturbation $\vert\delta p^{_{\rm NA}}_{_{T\gamma}}\vert$ 
(in red) has been plotted as a function of the number of $e$-folds~$N$ 
for the two cases of $\Gamma$, as in the previous two figures.  
For computational reasons, we have chosen a very small scale mode 
with wavenumber $k=10^{16}\;{\rm Mpc^{-1}}$ that exits the Hubble 
radius during inflation around $58$ $e$-folds or so.
Also, we have normalized the values of all the three quantities to 
be unity at Hubble exit.
Clearly, while the Bardeen potential remains nearly a constant 
across the transition, both the intrinsic entropy and the 
isocurvature perturbations decay in almost the same fashion before 
as well as after the transition.
The dashed lines in black and cyan indicate the $e^{-(2\,N)}$ and 
$e^{-(80\, N)}$ behavior, respectively.
Interestingly, the non-adiabatic pressure perturbations die down 
extraordinarily rapidly after reheating is complete.
It is these decay which ensure that the amplitude of the total 
curvature perturbation remains unaffected.
Note that these plots correspond to the potential $V_{1}(T)$ with 
values for the various parameters and the initial conditions that 
we have used in the last two figures.
As in the earlier figures, we find that a very similar behavior occurs
for the potential $V_{2}(T)$.}
\end{center}
\end{figure}

In Fig.~\ref{fig:enapp}, we have plotted the amplitudes of the Bardeen 
potential (i.e. $\vert\Phi\vert$) and the non-adiabatic pressure 
perturbations (viz. $\vert\delta p_{_{T}}^{_{\rm NA}}\vert$ and 
$\vert\delta p_{_{T\gamma}}^{_{\rm NA}}\vert$) as a function of the 
number of $e$-folds.
For convenience in numerical computation, we have chosen a very 
small scale mode (corresponding to $k=10^{16}\;{\rm Mpc^{-1}}$) 
that leaves the Hubble radius around $58$ $e$-folds or so 
during inflation\footnote{The amplitude of the non-adiabatic 
pressure perturbations corresponding to the cosmological scales 
(i.e. $10^{-4} \lesssim k \lesssim 1\;{\rm Mpc^{-1}}$) prove to 
be very small when they are well outside the Hubble radius.
As a result, evolving these reliably until the transition to the 
radiation dominated epoch requires considerable numerical accuracy, 
which is difficult to achieve.
For this reason, to illustrate the evolution of the non-adiabatic 
pressure perturbations, we choose to work with a very small scale 
mode that leaves the Hubble radius close to the end of inflation.
Since the inflationary epoch is of the slow roll type for all the 
$e$-folds, the non-adiabatic pressure perturbations associated with 
the cosmological scales can be expected to behave in a fashion 
similar to that associated with such a small scale mode.}.
The following points are evident from these two plots.
To begin with, barring a very small change in its amplitude---which 
is expected during the transition from inflation to the radiation 
dominated epoch---the Bardeen potential $\Phi$ remains a constant at 
super-Hubble scales.
Secondly, it is clear from the plots that the intrinsic entropy 
perturbation associated with the tachyon 
$\vert\delta p_{_{T}}^{_{\rm NA}}\vert$ decays as ${\rm e}^{-(2N)}$
at super-Hubble scales during inflation, a result that is well known 
in the literature (see, for instance, Refs.~\cite{leach-2001,rajeev-2007}).
Lastly, and interestingly, we find that the isocurvature perturbation 
$\vert\delta p^{_{\rm NA}}_{_{T\gamma}}\vert$ behaves in almost the 
same fashion as the intrinsic entropy perturbation both before and 
after the transition, a feature that has been noticed 
earlier~\cite{bartolo-2004}.
While they both behave as ${\rm e}^{-(2N)}$ at super-Hubble scales 
before the transition, they die down {\it extremely}\/ rapidly (roughly 
as ${\rm e}^{-(80N)}$!) during the radiation dominated epoch.
It is these behavior which ensure that the amplitude of the total 
curvature perturbation remains unaffected.

%%%%%%%%%%%%%%%%%%%%%%%%%%%%%%%%%%%%%%%%%%%%%%%%%%%%%%%%%%%%%%%%%%%%%%%%%%%%%%%

\section{Summary and discussion}\label{sec:d}

In this work, we have studied the evolution of perturbations in an 
interacting system consisting of a tachyon and radiation.
Treating the tachyon as an inflaton, we have investigated the effects
of reheating---i.e. the perturbative transfer of energy from the 
tachyon to radiation---on the large scale curvature perturbations. 
We have shown that the transition does not alter the amplitude of the 
total curvature perturbation of the system when the rate describing the 
decay of the inflaton into radiation is either a constant or a function 
of the tachyon.
We have also illustrated that, before the transition to the radiation 
dominated epoch, the relative non-adiabatic pressure perturbation 
between the tachyon and radiation decays in a fashion very similar to 
that of the intrinsic entropy perturbation associated with the tachyon.
Moreover, we have shown that, after the transition, the relative 
non-adiabatic pressure perturbation between the tachyon and radiation
dies down extremely rapidly during the early stages of the radiation 
dominated epoch. It is these behavior which ensure that the amplitude 
of the curvature perturbations remain unaffected during reheating.

It may be considered an overkill to integrate the second order 
differential equations rather than work with the much simpler 
first order equations~\cite{matarrese-2003,malik-2003,malik-2001}. 
Our motivations were threefold. 
Importantly, our effort allowed us to cross-check certain previous 
results~\cite{matarrese-2003}. 
Also, our aim is to later study the effects of deviations from slow 
roll inflation as well as the effects of the modified dynamics 
close to the transition to the radiation dominated epoch on the small 
scale perturbations. 
As we had mentioned in the introduction, these effects can have 
interesting implications for the number density of primordial black 
holes that are formed towards the end of inflation~\cite{pbhf}.
Moreover, in possibilities such as the warm inflationary scenario, 
the decay rate, in addition to depending on the inflaton, can also
depend on the temperature (see, for example, 
Refs.~\cite{wti,wi,deshamukhya-2009}).
In such a case, the transfer of the energy from the inflaton to 
radiation may affect the amplitude and possibly the spectral index 
as well~\cite{cerioni-2008}.

We would like to conclude by commenting on our choice for $\Gamma(T)$.
As we had pointed out earlier, our choices were motivated by 
convenience in numerical evolution rather than physics.
Obviously, better motivated functional forms of these quantities need 
to be investigated.
We are currently exploring such issues.
 
%%%%%%%%%%%%%%%%%%%%%%%%%%%%%%%%%%%%%%%%%%%%%%%%%%%%%%%%%%%%%%%%%%%%%%%%%%%%%%%

\section*{Acknowledgments}

We would like to thank Raul Abramo, Sudipta Das, Ruth Durrer, 
Lev Kofman, Andrew Liddle, Misao Sasaki, Ashoke Sen and David Wands 
for valuable discussions. 
We would also like to acknowledge the use of the cluster computing 
facilities at the Harish-Chandra Research Institute, Allahabad, India,
and at the Korea Institute for Advanced Study (KIAS), Seoul, Korea.
R.K.J also wishes to thank KIAS for hospitality, where part of this work 
was carried out.

%%%%%%%%%%%%%%%%%%%%%%%%%%%%%%%%%%%%%%%%%%%%%%%%%%%%%%%%%%%%%%%%%%%%%%%%%%%%%%%

\appendix

%%%%%%%%%%%%%%%%%%%%%%%%%%%%%%%%%%%%%%%%%%%%%%%%%%%%%%%%%%%%%%%%%%%%%%%%%%%%%%%

\section{The case of the canonical scalar field and a perfect fluid}

In this appendix, we shall illustrate the corresponding effects for 
the case of the canonical scalar field model that was considered 
recently in the literature~\cite{dimarco-2007}.
We shall rapidly summarize the essential equations describing the 
background evolution and the perturbations, and present the results.

%%%%%%%%%%%%%%%%%%%%%%%%%%%%%%%%%%%%%%%%%%%%%%%%%%%%%%%%%%%%%%%%%%%%%%%%%%%%%%%
\subsection{Background equations}

In the case of a canonical scalar field, say, $\phi$, that is 
interacting with a perfect fluid, if one assumes that $Q_{_{F}}
=(\Gamma\, {\dot \phi}^2)$, then~\eq{eq:ceic} that describes 
the conservation of the energy density of the fluid is given by
\beq
{\dot \rho}_{_{F}} + 3\, H \l(1 + w_{_{F}}\r)\, \rho_{_{F}} 
= \Gamma\; {\dot \phi}^2.
\eeq
Then, \eq{eq:c} implies that $Q_{_{\phi}}=-Q_{_{F}}
= -(\Gamma\; \dot{\phi}^2)$ and, hence, the continuity equation 
governing the energy density $\rho_{_{\phi}}$ of the scalar field 
reduces to
\beq
{\dot \rho}_{_{\phi}} + 3\,H \l(\rho_{_{\phi}} + p_{_{\phi}}\r)\, 
= -\Gamma\, {\dot \phi}^2,\label{eq:cephi}
\eeq
where $p_{_{\phi}}$ denotes the pressure of the field. 
The energy density and pressure associated with the canonical scalar 
field are given by 
\beq
\rho_{_{\phi}} = \frac{1}{2} {\dot \phi}^2 + V(\phi)
\qquad{\rm and}\qquad
p_{_{\phi}} = \frac{1}{2} {\dot \phi}^2 - V(\phi).
\eeq
Upon using these expressions in the continuity equation~(\ref{eq:cephi}), 
we arrive at the following equation of motion governing the scalar 
field~\cite{rh,dimarco-2007,cerioni-2008}:
\beq
{\ddot \phi} + 3\, H\; {\dot \phi}+ \Gamma\; {\dot \phi}
+V_{\phi}=0,\label{eq:emcsf}
\eeq
where $V_{_{\phi}}\equiv \l({\rm d}V/{\rm d}\phi\r)$. 

%%%%%%%%%%%%%%%%%%%%%%%%%%%%%%%%%%%%%%%%%%%%%%%%%%%%%%%%%%%%%%%%%%%%%%%%%%%%%%%

\subsection{Equations governing the scalar perturbations}

Upon using the expressions for the perturbed energy density, momentum 
flux and pressure of the scalar field in \eq{eq:pece}, we can 
arrive at the equation of motion for the perturbation in the scalar 
field, say, $\delta\phi$, exactly as we did for the perturbation in the 
tachyon.
In the UCG, we find that $\delta\phi$ satisfies the differential equation
\beq
{\ddot {\delta\phi}} + \l(3\,H + \Gamma\r)\,{\dot {\delta\phi}}
+ \l[V_{_{\phi\phi}} - \l(\frac{1}{a^2}\r)\,\nabla^{2}\r]\,
{\delta \phi}
- {\dot \phi}\,{\dot A} 
+ \l(2\, V_{_{\phi}}+ \Gamma\, {\dot \phi}\r)\, A
-\l(\frac{\dot \phi}{a}\r)\, (\nabla^{2}\, B)
+ {\dot \phi}\, \delta\Gamma=0,\label{eq:deltaphi}
\eeq
where $V_{_{\phi\phi}}=\l({\rm d}^2\,V/{\rm d}\phi^2\r)$ and, as 
before, $\delta\Gamma$ denotes the first order perturbation in the 
decay rate. 
Again, on using \eq{eq:pece}, we find that the equation 
describing the conservation of the perturbed energy density of the
fluid ${\delta \rho_{_{F}}}$ can be written as
\beq
\dot{\delta \rho_{_{F}}} + 3\,H\,(1 + w_{_{F}})\, 
\delta \rho_{_{F}}
+ \l(\frac{1}{a^2}\r)\nabla^{2} {\psi_{_{F}}}
-\, 2\, \Gamma\, {\dot \phi}\, {\dot {\delta\phi}}
+ \Gamma\, {\dot\phi}^2\,A
-\l[\frac{(1+w_{_{F}})\,\rho_{_{F}}}{a}\r]\, \l(\nabla^2 B\r)
-\dot\phi^2\, \delta\Gamma = 0.
\eeq
Also, from the first order Einstein's equations~(\ref{eq:foee0i})
and~(\ref{eq:foeeii}), we can arrive at the following differential 
equation for the quantity $\psi_{_{F}}$ that describes the spatial
velocity of the fluid:
\beq
{\dot \psi}_{_{F}}+3\, H\, \psi_{_{F}} + w_{_{F}}\, 
\delta \rho_{_{F}} +\l(1+w_{_{F}}\r)\, \rho_{_{F}}\, A
+ \Gamma\, {\dot \phi}\,\delta\phi = 0.
\eeq

The perturbation variables that we shall evolve numerically are
${\cal Q}_{_{\phi}}=\delta\phi$ and the quantity ${\cal Q}_{_{F}}$ 
that we had introduced earlier [cf. \eq{eq:calQF}].
As in the case of the tachyon and the perfect fluid, using the 
background equations, the first order Einstein's equations for 
the system and the energy conservation equations for the scalar 
field and the fluid, we can arrive at the coupled second order 
differential equations~(\ref{eq:ceqtqf}) for ${\cal Q}_{_{\phi}}$ 
and ${\cal Q}_{_{F}}$, with $(\alpha,\beta)
=(\phi,F)$~\cite{dimarco-2007}. 
For the case wherein the decay rate $\Gamma$ is a constant,
the coefficients ${\cal F}_{_{\alpha \beta}}$ and 
${\cal G}_{_{\alpha \beta}}$ are now given by
\br
{\cal F}_{_{\phi \phi}} 
&=& \l(3\,H + \Gamma\r),\\
%%%%%%%%%%%%%%%%%%%%%%%%%%%%%%%%
{\cal F}_{_{\phi F}}
&=& \l(\frac{4\, \pi\, G}{H}\r)
\l(\frac{w_{_{F}}-1}{w_{_{F}}}\r)\; {\dot \phi}\;
\sqrt{(1+w_{_{F}})\, \rho_{_{F}}}\,,\\
%%%%%%%%%%%%%%%%%%%%%%%%%%%%%%%%
{\cal F}_{_{F \phi}}
&=& \l(\frac{4\, \pi\, G}{H}\r)\,{\dot \phi}
\l(1-w_{_{F}}\r)\, \sqrt{(1+w_{_{F}})\, \rho_{_{F}}}
+ \l(1+2\,w_{_{F}}\r)\,
\l(\frac{\Gamma\, {\dot \phi}}{\sqrt{\l(1+w_{_{F}}\r)\, 
\rho_{_{F}}}}\r),\\
%%%%%%%%%%%%%%%%%%%%%%%%%%%%%%%%
{\cal F}_{_{FF}} 
&=& 3\,H + \l(\frac{\Gamma\, {\dot \phi}^2}{\rho_{_{F}}}\r),\\
%%%%%%%%%%%%%%%%%%%%%%%%%%%%%%%%
{\cal G}_{_{\phi \phi}}
&=& V_{_{\phi \phi}} -\l(\frac{1}{a^2}\r)\,\nabla^{2}
+ \l(\frac{4\, \pi\, G}{H}\r)
\l[4\,V_{_{\phi}}\, {\dot \phi} 
+2\; \l(3\,H+\Gamma\r)\, {\dot \phi}^{2}
-\l(\frac{\Gamma\, {\dot \phi}^{2}}{\rho_{_{F}}}\r)\r]\nn\\
&-&\l(\frac{4\, \pi\, G}{H}\r)^2 {\dot \phi}^2
\l[2\,{\dot \phi}^2 
+ \l(\frac{\l(1+w_{_{F}}\r)^2}{w_{_{F}}}\r)\,\rho_{_{F}}\r],\\
%%%%%%%%%%%%%%%%%%%%%%%%%%%%%%%%
{\cal G}_{_{\phi F}}
&=& \l(\frac{4\, \pi\, G}{H}\r)
\l[-\l(\frac{3\,H\, {\dot \phi}}{2}\r)\,
\l(\frac{\l(1+w_{_{F}}\r)^2}{w_{_{F}}}\r)
+ \l(\frac{1}{2\,\rho_{_{F}}}\r) \l(\frac{w_{_{F}}-1}{w_{_{F}}}\r)
\Gamma\, {\dot \phi}^3 - 2\, V_{_{\phi}} - \Gamma\, {\dot \phi}\r]
\sqrt{\l(1+w_{_{F}}\r)\,\rho_{_{F}}}\nn\\
&+& \l(\frac{4\, \pi\, G}{H}\r)^2 {\dot \phi}
\l[2\,{\dot \phi}^2 
+ \l(\frac{\l(1+w_{_{F}}\r)^2}{w_{_{F}}}\r)\,\rho_{_{F}}\r]\,
\sqrt{\l(1+w_{_{F}}\r)\,\rho_{_{F}}}\,,\\
%%%%%%%%%%%%%%%%%%%%%%%%%%%%%%%%
{\cal G}_{_{F \phi}}
&=& -\l(\!\frac{4\, \pi\, G}{H}\!\r)
\l[V_{_{\phi}}\, \l(1+w_{_{F}}\r)^2\,\rho_{_{F}}\,
+3\,H\, V_{_{\phi}}\, {\dot \phi}\, \l(1+w_{_{F}}\r)^2\, \rho_{_{F}}
-\Gamma\, {\dot \phi}^3\r]
\l(\!\frac{1}{\sqrt{\l(1+w_{_{F}}\r)\, \rho_{_{F}}}}\!\r)\nn\\
&+& \l(\frac{4\, \pi\, G}{H}\r)^2
\l({\dot \phi}^2 + 2\,\rho_{_{F}}\r)
\l(\frac{\l(1+w_{_{F}}^2\r)\,\rho_{_{F}} 
{\dot \phi}\,}{\sqrt{\l(1+w_{_{F}}\r)\,\rho_{_{F}}}}\r)
-\l(\frac{\Gamma}{\sqrt{\l(1+w_{_{F}}\r)\,\rho_{_{F}}}}\r)
\l(V_{_{\phi}}-3\, H \,w_{_{F}}\, {\dot \phi} 
+ \Gamma\, {\dot \phi}\r),\\
%%%%%%%%%%%%%%%%%%%%%%%%%%%%%%%%
{\cal G}_{_{FF}}
&=& -\l(\frac{w_{_{F}}}{a^2}\r)\, \nabla^{2} 
+ \l(\frac{9\, H^2}{4}\r)\, \l(1-w_{_{F}}^2\r)\nn\\
&+& \l(\frac{4\, \pi\, G}{H}\r)
\l[\l(\frac{3\,H}{2}\r) \l(1+w_{_{F}}\r) \l(1+3\,w_{_{F}}\r)\, \rho_{_{F}}
+ \l(\frac{3\,H}{2}\r) {\dot \phi}^2 \l(w_{_{F}}-1\r) -\Gamma\, {\dot \phi}^2\r]\nn\\
&-& \l(\frac{4\, \pi\, G}{H}\r)^{2}\!
\l[\l(1+w_{_{F}}\r)^2 \rho_{_{F}} {\dot \phi}^2
+ 2\, \l(\l(1+w_{_{F}}\r) \rho_{_{F}}\r)^2\r]
- \l(\frac{\Gamma\, {\dot \phi}}{\rho_{_{F}}}\r)\!
\l[V_{_{\phi}}+\Gamma\, {\dot \phi} 
- \l(\frac{3\,H}{2}\r) w_{_{F}} {\dot \phi}
+ \l(\frac{\Gamma\, {\dot \phi}^3}{4\,\rho_{_{F}}}\r)\r].
\er

For the case wherein the decay rate is a function of the scalar field, 
the coefficients ${\cal G}_{_{\phi \phi}}$ and ${\cal G}_{_{F \phi}}$ 
are modified to 
\br
{\cal G}_{_{\phi \phi}} 
& \to & {\cal G}_{_{\phi \phi}}
+ {\dot \phi}\;\Gamma_{_{\phi}},\\
{\cal G}_{_{F \phi}} 
& \to & {\cal G}_{_{F \phi}}
+ \l(\frac{w_{_{F}}}{\sqrt{\l(1+w_{_{F}}\r)\, \rho_{_{F}}}}\r)\, 
{\dot \phi}^2\; \Gamma_{_{\phi}},
\er
where $\Gamma_{_{\phi}}\equiv \l({\rm d}\,\Gamma/{\rm d}\phi\r)$. 
The other coefficients remain unchanged.

%%%%%%%%%%%%%%%%%%%%%%%%%%%%%%%%%%%%%%%%%%%%%%%%%%%%%%%%%%%%%%%%%%%%%%%%%%%%

\subsection{Results}

\begin{figure}[!t]
\begin{center}
\vskip 15pt\hskip 15pt
\resizebox{200pt}{130pt}{\includegraphics{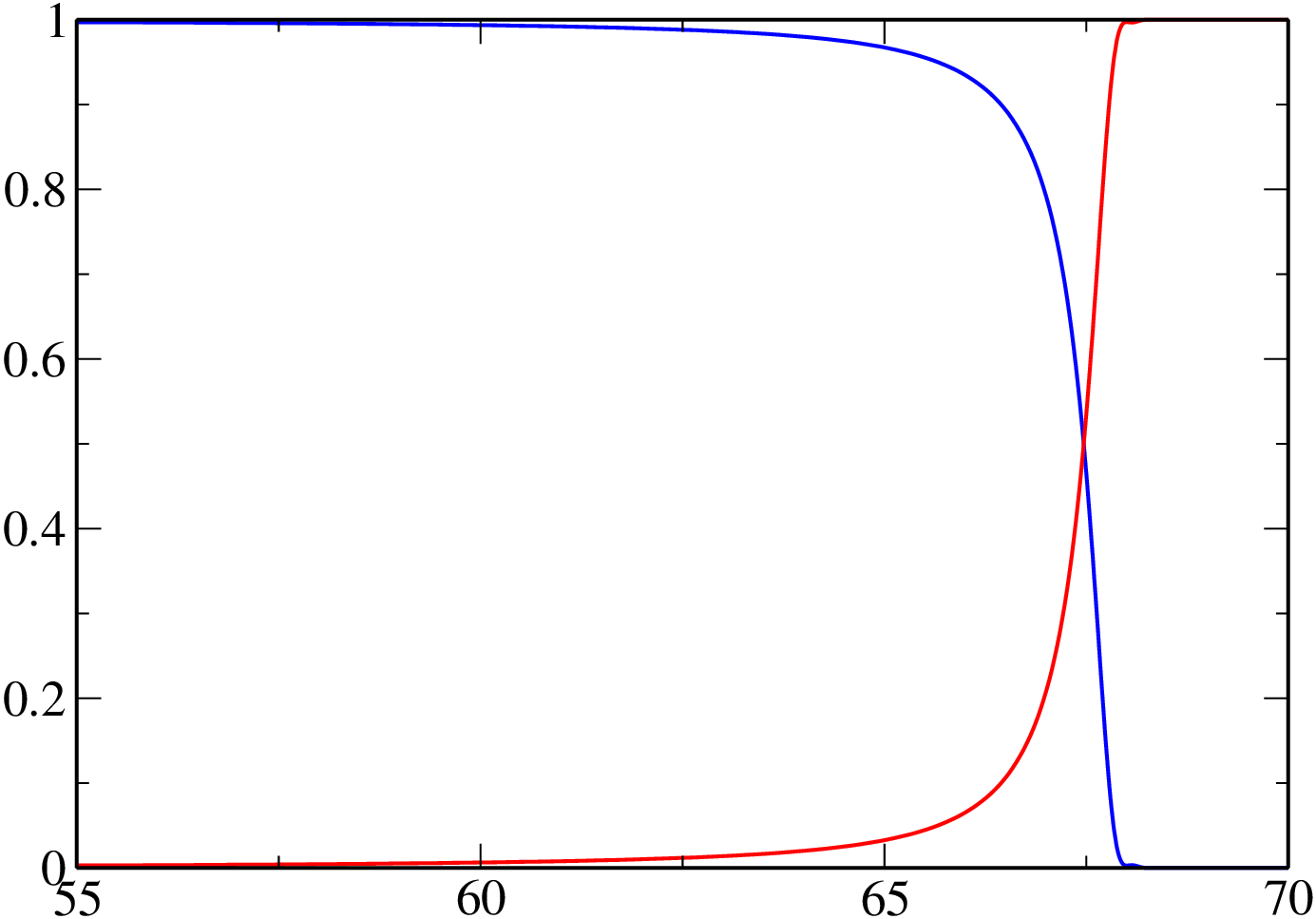}}
\hskip 25pt
\resizebox{200pt}{130pt}{\includegraphics{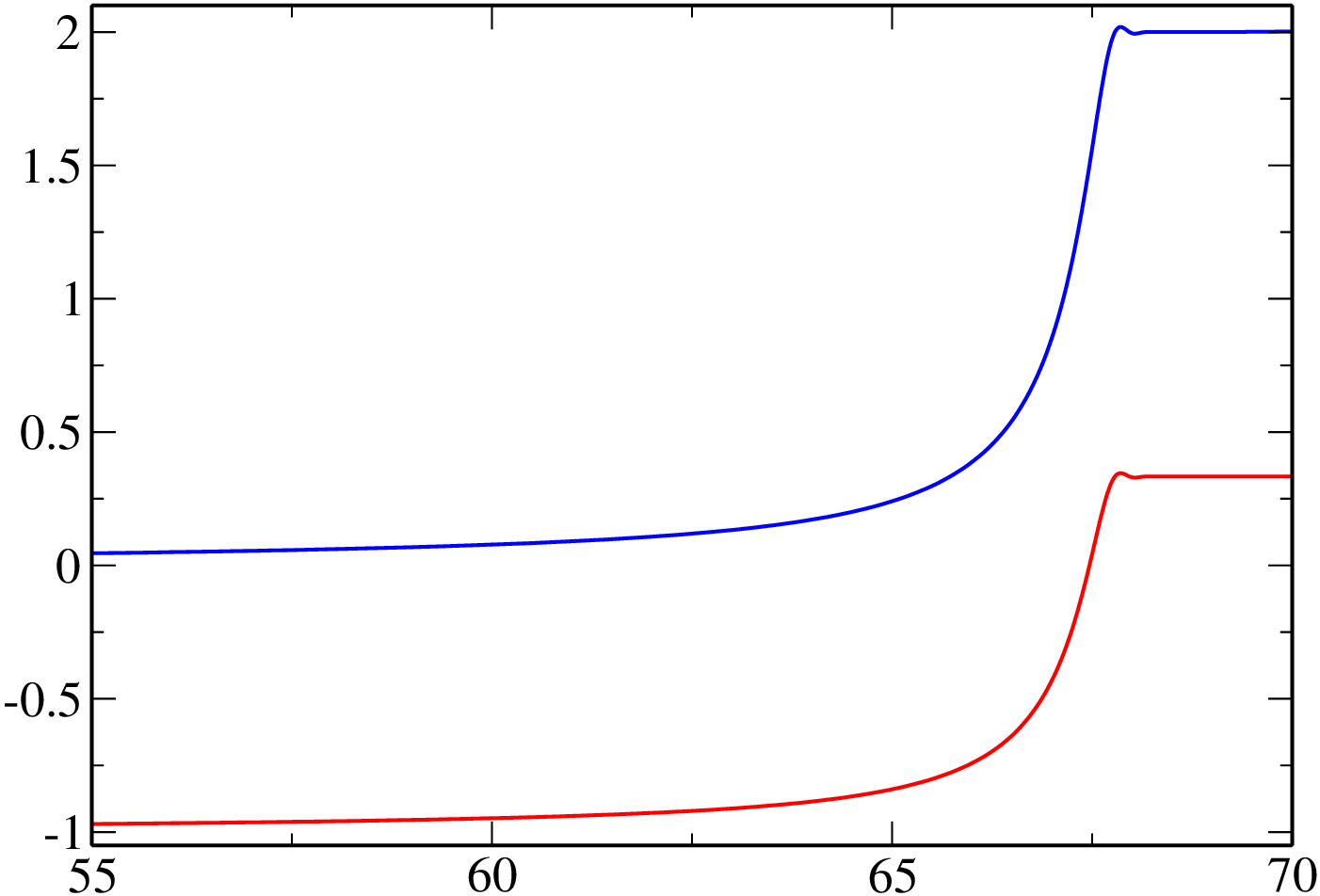}}
\vskip -100 true pt 
\hskip -220 true pt $\Omega_{_{\phi}}, \Omega_{_{\gamma}}$
\hskip 210 true pt $\epsilon,w$
\vskip 105 pt\hskip 15pt
\resizebox{200pt}{130pt}{\includegraphics{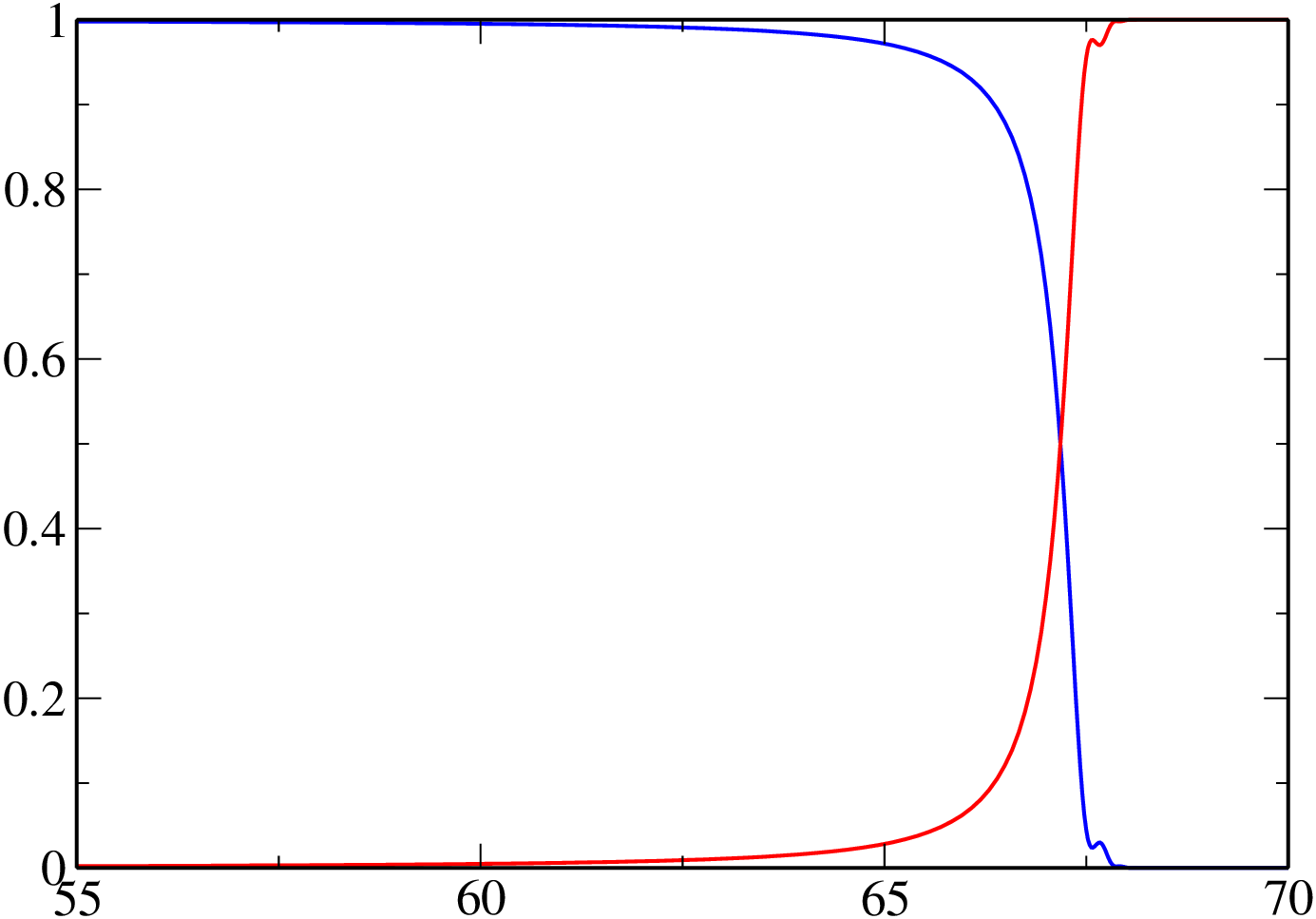}}
\hskip 25pt
\resizebox{200pt}{130pt}{\includegraphics{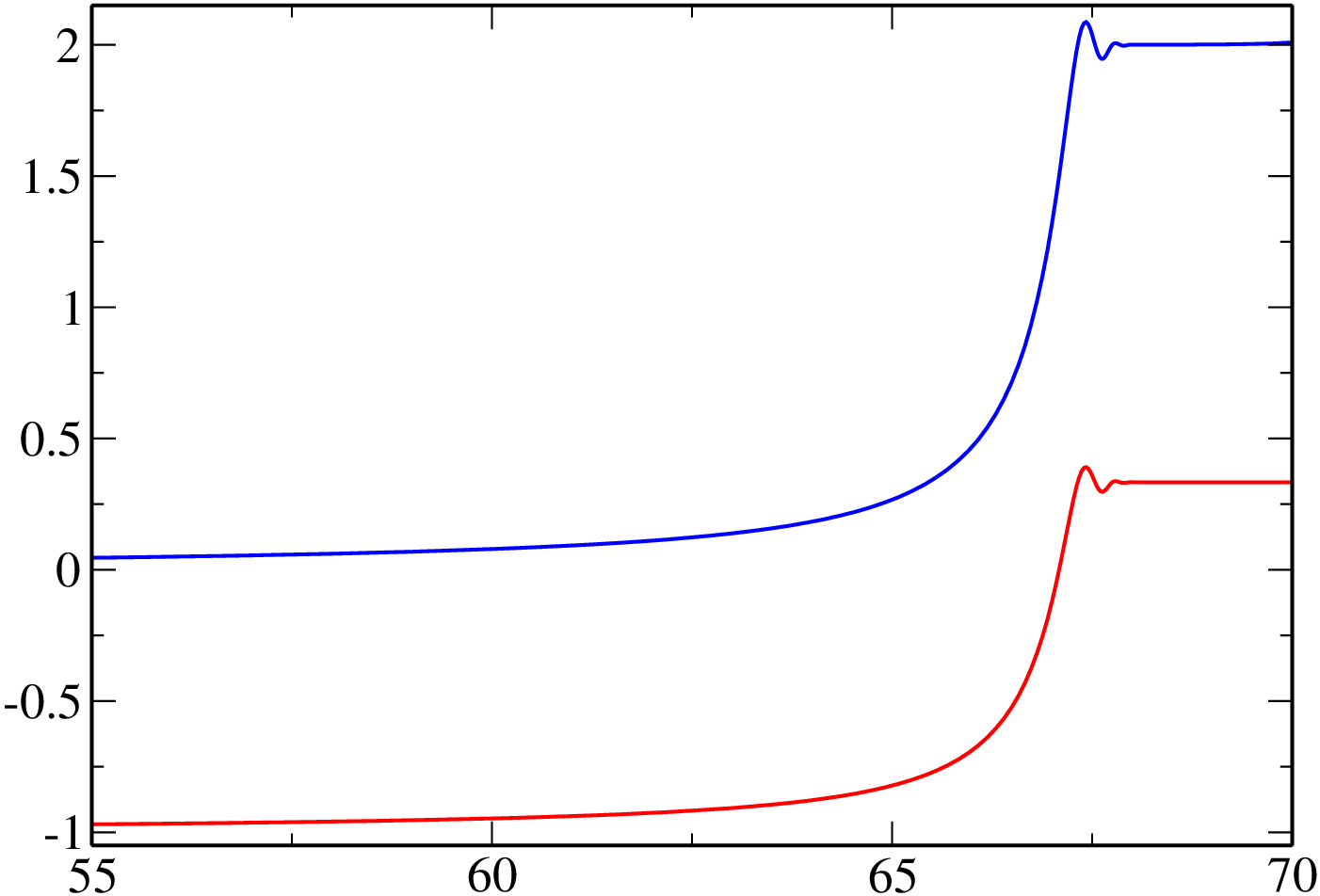}}
\vskip -100 true pt 
\hskip -220 true pt $\Omega_{_{\phi}}, \Omega_{_{\gamma}}$
\hskip 210 true pt $\epsilon,w$
\vskip 95 pt
\hskip 25 true pt $N$ \hskip 215 true pt $N$
\vskip 5pt
\caption{\label{fig:eewcsf}In the left column, the evolution of the 
quantities $\Omega_{_{\phi}}$ (in blue) and $\Omega_{_{\gamma}}$ (in
red) has been plotted as a function of the number of $e$-folds~$N$ for 
the two types of $\Gamma$, as discussed in the earlier figures. 
In the right column, we have plotted the evolution of the first Hubble 
slow roll parameter~$\epsilon$ (in blue) and the equation of state 
parameter~$w$ of the entire system (in red).
We have considered the popular quadratic potential to describe 
the scalar field, and we have worked with suitable parameters 
and initial condition to arrive at the required behavior. 
These plots clearly illustrate the transfer of energy from the 
inflaton to radiation. 
We should point out that the oscillations in the plots arise due
the oscillations by the field at the bottom of the potential, a
feature that does not arise in the case of the tachyon.}
\end{center}
\end{figure}

\begin{figure}[!htb]
\begin{center}
\vskip 15pt\hskip 15pt
\resizebox{200pt}{130pt}{\includegraphics{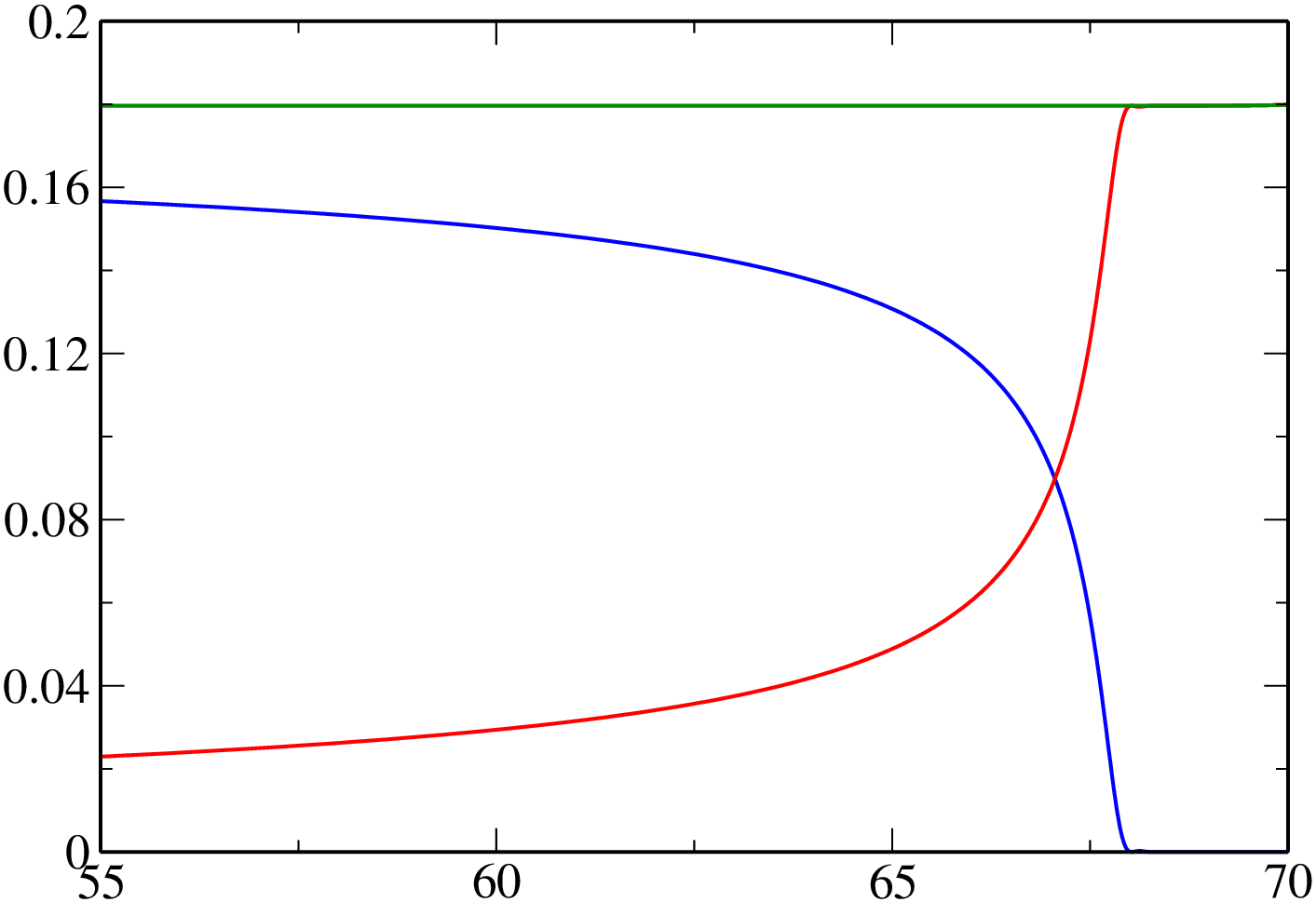}}
\hskip 25pt
\resizebox{200pt}{130pt}{\includegraphics{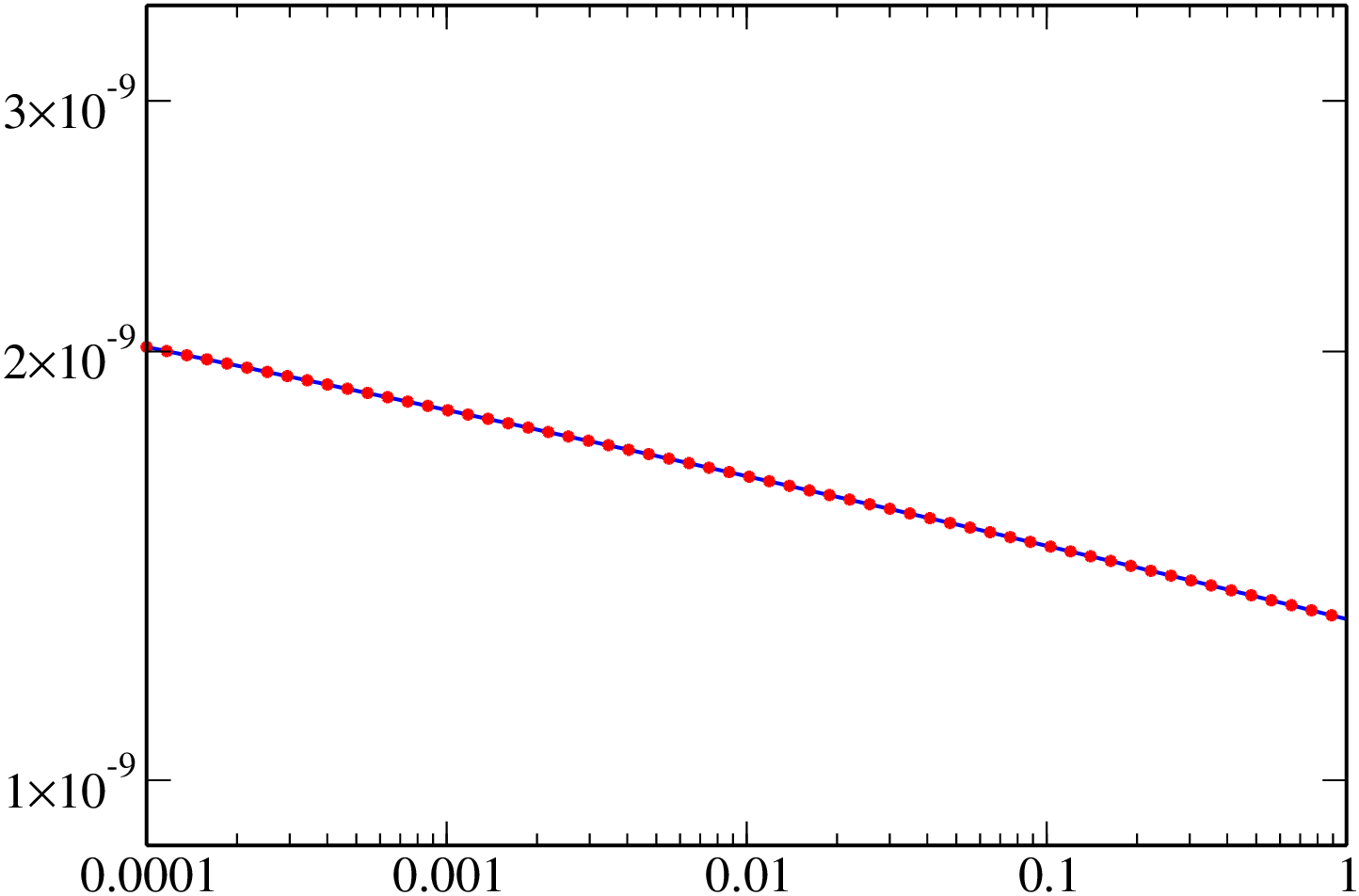}}
\vskip -105 true pt 
\hskip -245 true pt $\vert{\cal R}_{_{\phi}}\vert, 
\vert{\cal R}_{_{\gamma}}\vert, \vert{\cal R}\vert$
\hskip 210 true pt ${\cal P}_{_{\rm S}}(k)$
\vskip 110 pt\hskip 15pt
\resizebox{200pt}{130pt}{\includegraphics{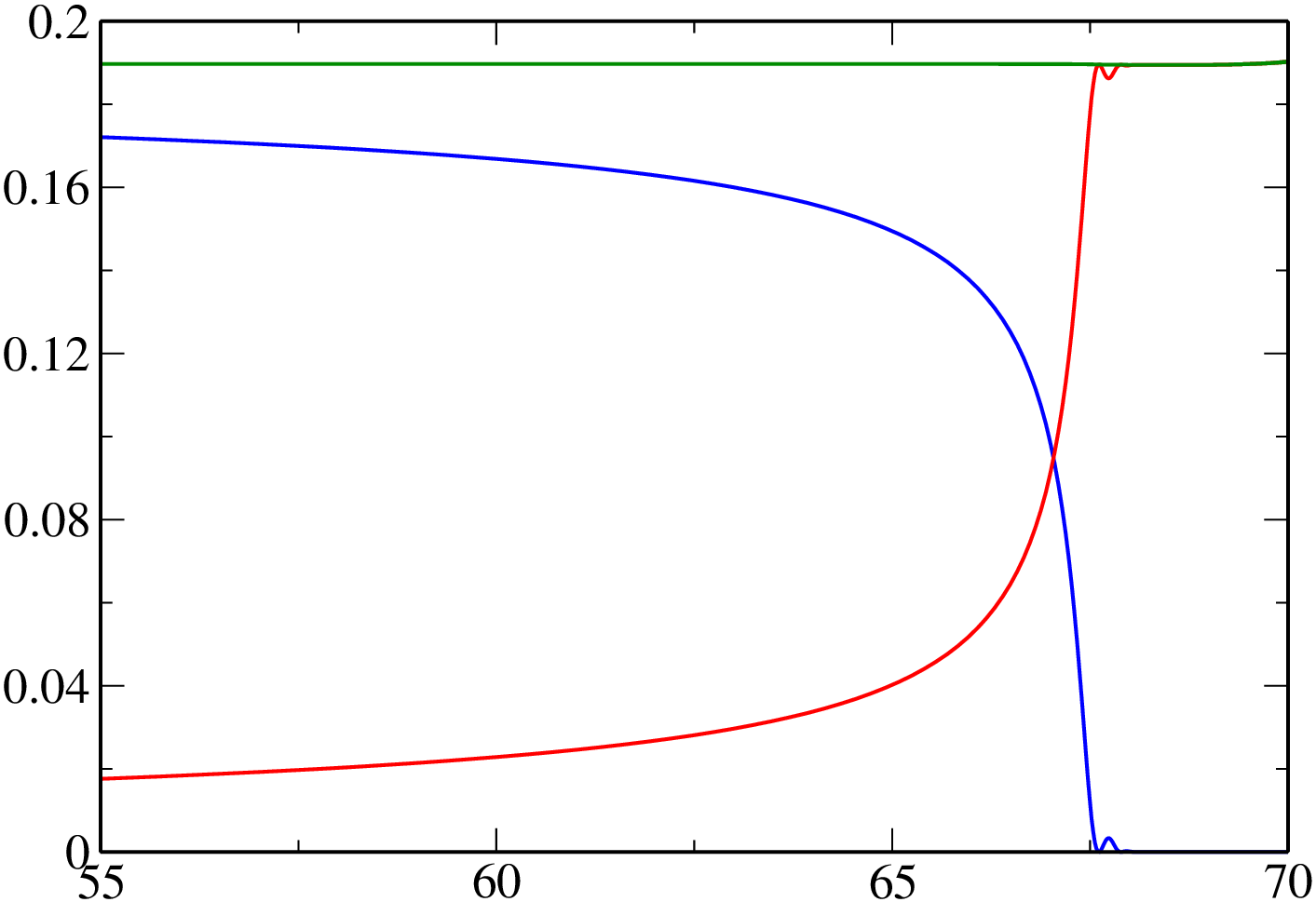}}
\hskip 25pt
\resizebox{200pt}{130pt}{\includegraphics{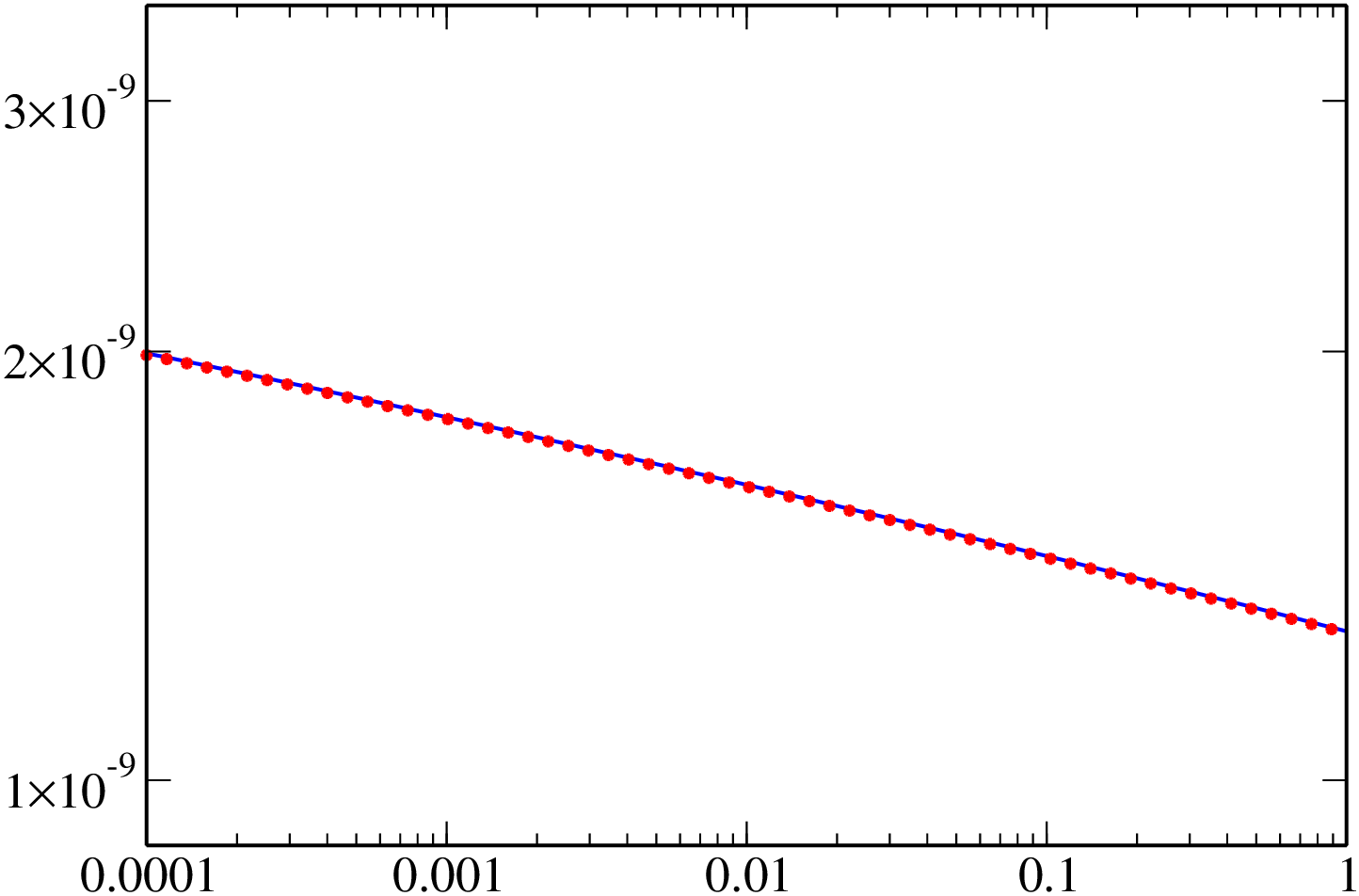}}
\vskip -105 true pt 
\hskip -245 true pt $\vert{\cal R}_{_{\phi}}\vert, 
\vert{\cal R}_{_{\gamma}}\vert, \vert{\cal R}\vert$
\hskip 210 true pt ${\cal P}_{_{\rm S}}(k)$
\vskip 95 pt
\hskip 25 true pt $N$ \hskip 215 true pt $k$
\vskip 5pt
\caption{\label{fig:epscsf}In the left column, the evolution of
the amplitudes of the curvature perturbations associated with the 
scalar field ($\vert{\cal R}_{_{\phi}}\vert$, in blue), radiation 
($\vert{\cal R}_{_{\gamma}}\vert$, in red) and the total curvature 
perturbation of the entire system ($\vert{\cal R}\vert$, in green) 
has been plotted as a function of the number of $e$-folds~$N$ for 
the two types of the decay rate.
For illustration, we have again chosen a typical cosmological mode 
with wavenumber $k=0.01\;{\rm Mpc^{-1}}$.
In the right column, as before, we have plotted the scalar power 
spectrum before (in blue) and after (in red) the transition.
It is evident that, as in the case of the tachyon, the amplitude 
of the nearly scale invariant power spectrum remains unaffected for 
both types of the decay rate.}
\end{center}
\end{figure}

In Fig.~\ref{fig:eewcsf}, we have illustrated the transition 
from inflation to the radiation dominated epoch by plotting 
the dimensionless energy density parameters $\Omega_{_{\phi}}$ 
and $\Omega_{_{\gamma}}$ for the two possible types of decay. 
We have also plotted the evolution of the first Hubble slow 
roll parameter $\epsilon$ and the equation of state parameter~$w$ 
for the entire system.
Note that we have considered the quadratic potential~$V(\phi)
=(m^2\, \phi^{2}/2)$, and have chosen suitable values for the
parameters and initial conditions to achieve the desired behavior.
And, in Fig.~\ref{fig:epscsf}, as in the case of the tachyon discussed 
in the text, we have plotted the evolution of the amplitudes of the
individual and the total curvature perturbations, as well as the 
scalar power spectrum before and after the transition.
It is evident from these two figures that the conclusions we had 
arrived at for the tachyon apply equally well to the case involving 
the canonical scalar field as well. 

%%%%%%%%%%%%%%%%%%%%%%%%%%%%%%%%%%%%%%%%%%%%%%%%%%%%%%%%%%%%%%%%%%%%%%%%%%%%%%%
%\section*{References}

%%%%%%%%%%%%%%%%%%%%%%%%%%%%%%%%%%%%%%%%%%%%%%%%%%%%%%%%%%%%%%%%%%%%%%%%%%%%%%%
\end{document}